\documentclass[twocolumn,reprint,prb,aps,showpacs,msmath,amsmath,amssymb]{revtex4-2}
\usepackage[colorlinks=true,urlcolor=blue,linkcolor=blue,citecolor=blue]{hyperref}
\usepackage{graphicx}
\usepackage[dvipsnames]{xcolor}
\usepackage{ulem}
\usepackage{cancel}
\begin{document}

% input definitions
%\input{rmmcoms}

%RMM-Oct. 1, 2021 - added shortcut commands for color - can change later by global replace
\newcommand{\CRD}{\color{red}}
\newcommand{\CBK}{\color{black}}
\newcommand{\CBLU}{\color{blue}}

% **************************************************

\title{Electronic Structure and Magnetism of the Triple-layered Ruthenate Sr$_4$Ru$_3$O$_{10}$ }

\author{G.~Gebreyesus}
\email{ghagoss@ug.edu.gh}
\affiliation{Department of Physics, School of Physical and Mathematical Sciences, College of Basic and Applied Sciences, University of Ghana, Ghana}

\author{Prosper Ngabonziza}
\email{p.ngabonziza@fkf.mpg.de}
\affiliation{Max Planck Institute for Solid State Research,
Heisenbergstraße 1, 70569 Stuttgart, Germany}
\affiliation{Department of Physics, University of Johannesburg, P.O. Box 524 Auckland Park 2006, Johannesburg, South Africa}

\author{Jonah Nagura}
\affiliation{Department of Physics, Federal University of Agriculture, Abeokuta, Nigeria}

\author{Nicola Seriani}
\affiliation{The Abdus Salam ICTP, I-34151, Trieste, Italy}

\author{Omololu Akin-Ojo}
\affiliation{East Africa Institute for Fundamental Research, University of Rwanda, Kigali, Rwanda}

\author{Richard M.~Martin}
  \email{rmartin@illinois.edu}
\affiliation{Department of Physics, University of Illinois at
         Urbana-Champaign, Urbana, IL 61801, USA}
\affiliation{Department of Applied Physics, Stanford University,
        Stanford, CA 94305, USA}

\date{\today}
%\maketitle

\begin{abstract}

We report electronic band structure calculations for Sr$_4$Ru$_3$O$_{10}$ which displays both ferromagnetic and metamagnetic behavior. The density functional calculations find the ground state to be ferromagnetic in agreement with experiment and we show that the resulting spin polarization has dramatic consequences for the electronic properties. The minority spin bands are mainly empty and disperse steeply upward at the Fermi energy, whereas the majority spin bands are full or nearly fully occupied and form narrow bands near the Fermi energy, which could be the electronic origin of the metamagnetism.
Inclusion of a Hubbard interaction U applied to the Ru $4d$ states has major effects on the narrow bands, which reveal the role of Coulomb interactions and correlated many-body physics. The results are in qualitative agreement with recent angle resolved photoemission spectroscopy (ARPES) experiments and show the need for a combined theoretical study and experimental ARPES investigation with better energy resolution to reveal the nature of the narrow bands close to the Fermi-level, which is critical for understanding the exotic magnetic properties observed in this material.
\CBK
%
%We report electronic band structure calculations for Sr$_4$Ru$_3$O$_{10}$ which displays both ferromagnetic and metamagnetic behavior. The density functional calculations find the ground state to be ferromagnetic in agreement with experiment and we find that the inclusion of Coulomb Hubbard interaction U applied to the Ru $4d$ states has dramatic effects on the Fermi surface, which reveal the role of Coulomb interactions and correlated many-body physics.  The minority spin bands are mainly empty with Fermi surfaces in the outer areas of the Brillouin zone away from the $\Gamma$ point with bands that disperse steeply upward.   The majority spin bands are full or nearly fully occupied and form narrow bands near the Fermi energy around the $\Gamma$ point, which could be the electronic origin of the metamagnetism.  The results are in qualitative agreement with recent angle resolved photoemission spectroscopy (ARPES) experiments and show the need for a combined theoretical study and experimental ARPES investigation with better energy resolution to reveal the nature of the narrow bands close to the Fermi-level, which is critical for understanding the exotic magnetic properties observed in this material.

\end{abstract}

\pacs{}  %CHECK!

\maketitle

\section{Introduction}
\label{sec:Introduction}
%Removed long paragraphs that were commented out in previous versions

Strontium ruthenate materials Sr$_{n+1}$Ru$_n$O$_{3n+1}$ form a unique class of layered compounds in  the Ruddlesden Popper (RP) series; because the number ($n=1,2,3,\cdots, \infty$) of the RuO$_6$ octahedra in the unit cell represents a critical parameter for obtaining distinct collective phenomena in the ground state, depending on the overlap between Ru $4d$ wave functions through adjacent octahedra~\cite{MMalvestuto_2011,CBergemann_2003,AMackenzie_2003}. The rich array of ground state properties in this series include superconductivity, magnetic order and metamagnetism, orbital order, heavy fermions, and other phenomena~\cite{MMalvestuto_2011,CBergemann_2003,AMackenzie_2003,SIkeda_1999,SAGrigera_2001,RABorzi_2007,PRivero_2017,MKCrawford_2002,FWeickert_2017,GKoster_2012}. These materials are also particular examples of the 4$d$ transition metal oxides in which electronic correlations are remarkably strong. The richness of different phenomena in Sr-based layered ruthenates  comes from  a competition between local and itinerant physics; which is an evidence of strong interplay between charge, spin, orbital, and lattice degrees of
freedom~\cite{GCao_2003,ZQMao_2006,GCao_1997,MMalvestuto_2011}.

Among the strontium ruthenates in the RP series, the single layer Sr$_2$RuO$_4$ has been studies extensively, mostly because it is a  superconductor with an unconventional order~\cite{CBergemann_2003,AMackenzie_2003,KIshida_1998,AMackenzie_1998,GMLuke_1998,YMaeno_2001,SIIkeda_2000,Haverkort,Veenstra}; and it is the first example of a non-cuprate layered perovskite superconductor~\cite{YMaeno_1994}.   The electronic properties are due to bands associated with the Ru-O planes which form a square lattice with the same structure as the Cu-O planes in the high-temperature superconductors~\cite{ADamascelli_2000,ATamai_2017}. The other members of the layered strontium ruthenates are formed by stacks of similar layers.
The double layered Sr$_3$Ru$_2$O$_7$ has two layers of RuO$_6$ octahedra per unit cell and it is generally regarded as a paramagnetic that is very near a ferromagnetic instability~\cite{SIkeda_1999}. Due to its interesting diamagnetic features, quantum critical metamagnetism and nematic fluid behavior with heavy d-electron masses, the Sr$_3$Ru$_2$O$_7$ system has also been the subject of many theoretical and experimental investigations~\cite{SAGrigera_2001,RABorzi_2007,ATamai_2008,DJSingh_2001,MPAllan_2013,JLee_2009,SAGrigera_2004,BBinz_2004}.
At the other end of this series, the increased tendency toward magnetism culminates in SrRuO$_3$, which is a ferromagnetic metal with the cubic perovskite structure that can be viewed as the infinite layer limit of the series~\cite{GKoster_2012,HTDang_2015,LKlein_1996,PBAllen_1996}. Also, SrRuO$_3$ is a highly conductive metallic oxide with  a good lattice match to many oxide materials; thus it has been widely used as a conductive electrode in diverse oxide heterostructures~\cite{GKoster_2012,PNgabonziza_2021,YDing_2016,SVKumar_2014,SLee_2013}.

The focus of this paper is the triple-layer system Sr$_4$Ru$_3$O$_{10}$, which displays both anisotropic ferromagnetism and intriguing orbital-selective metamagnetic behavior~\cite{MKCrawford_2002,FWeickert_2017,YJJo_2007,GCao_2003,ZQMao_2006,DFobes_2010,YLiu_2010,YLiu_2010,ECarleschi_2014,WSchottenhamel_2016}.
%Initial works conducted on this compound indicated that it is a structurally distorted ferromagnet, featuring anisotropic magnetic and electronic  transport properties~\cite{MKCrawford_2002,GCao_2003,ZQMao_2006}.
The spontaneous ferromagnetic moment is perpendicular to the planes below the Curie temperature of $105$ K, with an additional transition at around $60$ K observed in magnetic susceptibility~\cite{YJJo_2007,GCao_2003,ZXu_2007,FWeickert_2017}. Depending of the sample quality, the reported saturated magnetic moments in Sr$_4$Ru$_3$O$_{10}$ single crystals range from $\sim 1.0$ to $\sim 1.7\mu_B/\text{Ru}$ oriented along the $c-$axis~\cite{MKCrawford_2002,GCao_2003,ZQMao_2006,MZhou_2005,FWeickert_2017, FForte_2019}. A spin-polarized neutron diffraction study~\cite{FForte_2019} has shown that the moments are largest on the central layer and are almost completely due to the spin with vanishingly small orbital contributions. For magnetic field applied along the $ab-$plane, there is a decrease in magnetization below the second transition at 60 K, which is accompanied with a metamagnetic transition
%in the magnetization versus applied magnetic field curve
for fields of about 2.5 T~\cite{MZhou_2005,DFobes_2010,YJJo_2007}. A double metamagnetic behavior was also reported in  Sr$_4$Ru$_3$O$_{10}$ with a second in-plane metamagnetic transition at a slightly larger applied field~\cite{ECarleschi_2014}.

Electronic band structure calculations of Sr$_2$RuO$_4$~\cite{TOguchi_1995,DJSingh_1995}, Sr$_3$Ru$_2$O$_7$~\cite{DJSingh_2001,ATamai_2008} and SrRuO$_3$~\cite{AHerklotz_2015,NMiao_2013} materials have been reported, and compared in detail to angle resolved photoemission (ARPES) experiments~\cite{ADamascelli_2003,ATamai_2017,ATamai_2008,MPAllan_2013,HRyu_2020,DEShai_2013}.  However, it is only recently that the band structure of the Sr$_4$Ru$_3$O$_{10}$ system has been studied by ARPES~\cite{PNgabonziza_2020},~\footnote{After this paper was submitted, we learned of previous ARPES measurements that are reported in talks in 2005 and 2008, but were never published. To our knowledge, they are consistent with the data in ~\cite{PNgabonziza_2020}. See Bulletin of the American Physical Society, 2005 March Meeting, {\it Three-Dimensional Band Structure of $Sr_{4}Ru_{3}O_{10}$}, F. Wang, J.W. Allen, J.D. Denlinger, X.N. Lin and Gang Cao~\url{https://meetings.aps.org/link/BAPS.2005.MAR.J25.8} and 2008 March Meeting, {\it Fermi Surface mapping of  $Sr_{4}Ru_{3}O_{10}$ using Angle Resolved Photoemission}, R.S. Singh, F. Wang, J.W. Allen, J.D. Denlinger, X.N. Lin and Gang Cao~\url{http://meetings.aps.org/link/BAPS.2008.MAR.X31.12}}.
	\CBK To our knowledge, there have been no electronic structure calculations for Sr$_4$Ru$_3$O$_{10}$ reported so far.
%; which, by comparing with recent ARPES data, would provide opportunity to understand better the nature of the electronic states and the origins magnetic behavior in Sr$_4$Ru$_3$O$_{10}$.
%Thus, to get a comprehensive description of the band structure and magnetism of this system, electronic structure calculations are highly desirable.
\CBK

Here, we report density functional calculations of the band structure and Fermi surfaces of Sr$_4$Ru$_3$O$_{10}$ that
reveal the large effects due to the magnetic properties of this system.
In agreement with experiment, the ground state of Sr$_4$Ru$_3$O$_{10}$ is found to be ferromagnetic, so that there are two sets of bands for the two spins.
%The extracted magnetic polarization is $\sim \textcolor{red}{ \text{ xxx }} \mu_B/\text {Ru}$. Contrary,
The minority spin bands have reduced occupation and Fermi surfaces that are mainly in the outer areas of the Brillouin zone (BZ) away from $\Gamma$ with  bands that disperse steeply upward.
%although there may also be hole-like bands dispersing downward around $\Gamma$.
In contrast, the majority spin bands are full or nearly fully occupied with the Fermi level near the top of the bands leading to narrow bands near the Fermi energy around the $\Gamma$ point.
As previously found for other Sr-based layered ruthenates~\cite{HLHuang_2020, Mario-1999, David-2020, Igor-2014},
we find that inclusion of a Hubbard $U$
has dramatic effects on states near the Fermi energy, signifying the role of Coulomb interactions and correlated many-body physics %in the  properties of  Sr$_4$Ru$_3$O$_{10}$ such as the anomalous metamagnetic behavior.
for explaining the exotic ground state properties, e.g., anomalous
metamagnetic behavior in this material.

The calculated bands are in general agreement with ARPES experiments~\cite{PNgabonziza_2020}, which find hole-like and electron-like pieces of the Fermi surfaces consistent with spin-polarized bands of a ferromagnetic system and very different from the calculated Fermi surfaces if the system is constrained to be non-spin-polarized.  However,
%we propose that
it will further require combined theoretical study and experimental ARPES investigation with higher energy resolution to sort out the details of the narrow bands near the Fermi energy.

The remainder of this paper is organized as follows. In Sec.~\ref{sec:Structure} we discuss the structure of the layered ruthenates with the focus on the three-layered ruthenate materials. Sec.~\ref{sec:Simplified-model} contains a discussion about the nature of the bands in these classes of materials, and presents results from a tight-binding model, which are very useful for understanding the DFT results and comparison with the experiment.
Sec.~\ref{sec:DFT- calcs} contains the technical details of our DFT calculations, and presents the band structures and Fermi surfaces for the spin-polarised and non-spin-polarised calculations.
In Sec.~\ref{sec:Exp-comparison} we discuss the comparison of the calculated bands and Fermi surfaces with the experimental data. Sec.~\ref{sec:Consequences} discusses the possible consequences for the interesting properties of Sr$_4$Ru$_3$O$_{10}$  and in Sec.~\ref{sec:Conclusions} are concluding remarks.

\section{Structures of the layered ruthenates}
\label{sec:Structure}

\begin{figure}
\begin{center}
\includegraphics[width=0.375\textwidth]{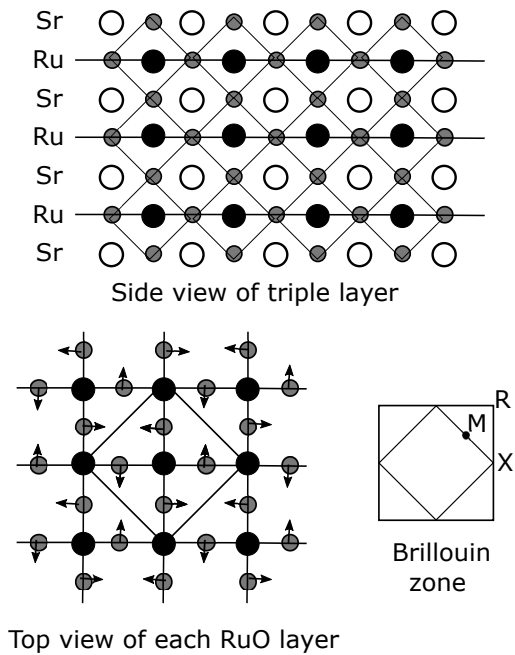}
\end{center}
\caption{The structure of Sr$_4$Ru$_3$O$_{10}$ is based on the triple layer shown at the top, which consists of 3 layers of Ru atoms surrounded by octahedra of O atoms indicated by the small grey circles. There are layers of Sr between the Ru layers and at the upper and lower sides of a triple layer.  The stacking of the layers in the crystal is described in the text but the main features of the electronic structure are determined by the bands of a single triple layer.  At bottom left is depicted a layer of Ru and O atoms. If the O atoms were exactly between the Ru atoms, the layer would have one Ru per cell; however, the octahedra are rotated as indicated by the arrows which doubles the primitive cell as shown by the square. Similarly the BZ is reduced by a factor of two and turned by 45 degrees as indicated at the lower right.}
% It is found that the rotation is opposite in the inner and outer layers\cite{MKCrawford_2002}.
 \label{fig:SrRuO-structure}
\end{figure}

The structures of the layered strontium ruthenates in the RP sequence Sr$_{n+1}$Ru$_n$O$_{3n+1}$, $n = 1,2, \ldots$  consist of Ru-O and Sr-O layers as illustrated in Fig.~\ref{fig:SrRuO-structure} for Sr$_4$Ru$_3$O$_{10}$.
At the top is the side view of a triple layer which can be considered as three layers of Ru and O atoms, with four Sr-O layers: two Sr-O layers between the Ru-O layers and two Sr-O layers at the top and bottom of each triple layer.
 The three-dimensional crystal is formed by stacking the triple layers with the outer Sr-O layers creating an effective insulating barrier between adjacent triple layers, so that the relevant electronic bands are essentially two-dimensional.  The calculations described in Sec.~\ref{sec:DFT- calcs} are for the experimentally determined three-dimensional structure~\cite{MKCrawford_2002}. However, the primary results are independent of the stacking and the properties are dominated by the two-dimensional bands of a single triple layer.
 This is verified by the calculations and is assumed in the interpretations of the ARPES experiments which are analyzed purely in terms of two-dimensional bands~\cite{PNgabonziza_2020}.

 At the bottom of Fig.~\ref{fig:SrRuO-structure} is a top view of a single Ru-O layer with the rotations of the octahedra indicated by the displacements of the oxygen atoms in the plane.  If there are no rotations, as in the single layer system Sr$_2$RuO$_4$, each Ru atom is equivalent and the primitive unit cell is a square containing one Ru atom.  In the double and triple layer systems, the rotations double the unit cell so that it is a square containing two Ru atoms and rotated by 45 degrees as shown in Fig.~\ref{fig:SrRuO-structure} (bottom left).  Correspondingly, the BZ is a square reduced by 1/2 and turned by 45 degrees, as depicted at the bottom right in Fig.~\ref{fig:SrRuO-structure}. In the triple layer, Sr$_4$Ru$_3$O$_{10}$, the rotations of the RuO$_{6}$ octahedra in the inner and outer layers are not equivalent; the RuO$_{6}$ octahedra in the inner layer rotated by an angle of $\approx$ 11 degrees about the c axis whereas the outer two layers rotated by very nearly half that amount in the opposite direction~\cite{MKCrawford_2002}.

% Because of the rotations of the octahedra the primitive cell in the plane contains 2 Ru atoms and is a square as shown.  In an idealized structure with no rotations of the octahedra all the Ru atoms in a layer are equivalent and the primitive unit cell is a square containing one Ru atom and rotated 45 degrees from the two-atom cell shown in Fig.~\ref{fig:SrRuO-structure}.

\section{Nature of the bands}
\label{sec:Simplified-model}

The main features of the band structures of the layered compounds in the RP series
%, Sr$_{n+1}$Ru$_n$O$_{3n+1}$, $n = 1,2, \ldots$,
can be understood starting from the band structure a single Ru-O layer shown at the bottom left in Fig.~\ref{fig:SrRuO-structure}.  In all cases, the Ru $d$ states involved are primarily the 3 $t_{2g}$ states d$_{xz}$, d$_{yz}$ and d$_{xy}$ where $x,y$ denote the directions in the plane of the layers and $z$ is perpendicular to the plane. The bands are formed by Ru-Ru hopping via the O $p$ states so that they are mixed Ru and O characters with the states near the Fermi energy composed primarily of Ru $d$ states.  There are also oxygen atoms (called apical oxygens) above and below the Ru atoms that have the effect of raising the energy of the d$_{xz}$, d$_{yz}$ relative to the d$_{xy}$.
%If we consider only the O p states,

\begin{figure}
\begin{center}
\includegraphics[width=0.5\textwidth]{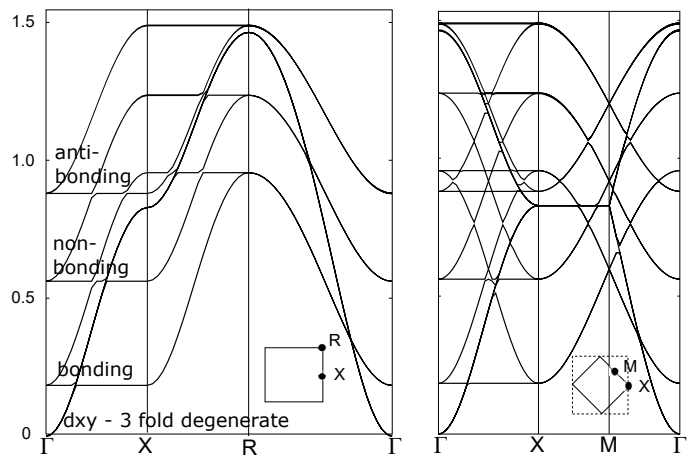}
\end{center}
\caption{Characteristic bands of the coupled Ru t$_{2g}$ and O p states found using the tight-binding model described in the text and the structure shown in Fig.~\ref{fig:SrRuO-structure}. Only the bands near the Fermi energy are shown. These have mainly Ru $d$ character; O states are at lower energy. At the left are bands of a triple layer which are like three copies of the single layer taking into account the splitting due to the hopping between neighboring planes in the triple layer. The figure at the right illustrates the effect of doubling the cell, which folds the bands into the smaller BZ and leads to states with maxima at the $\Gamma$ point. Note the similarity to the bands for the full DFT calculations shown in Figs. \ref{fig:bands-NM}, \ref{fig:Exp_Geo_bands_tot}, \ref{fig:Exp_Geo_bands_SOC-1to1} and \ref{fig:PBEsol-U2}.}
 \label{fig:tb-progression}
\end{figure}

It is very useful for understanding the DFT results and comparison with experiment to have simple models that capture the essential features.
The minimal model that captures the main features is a tight-binding model with only one p-d hopping matrix element t$_{pd\pi}$ and one relevant energy difference, $\Delta_{pd} = \varepsilon_d - \varepsilon_p$,
 which is the well-known model for the single layer Sr$_2$RuO$_4$  and used for other materials in the RP series~\cite{MMalvestuto_2011,CBergemann_2003,AMackenzie_2003}.
%For many purposes it is sufficient to models the bands near the Fermi energy as $t_{2g}$ states with determined by a single parameter, the effective hopping between nearest neighbors Ru atoms $t_{eff} = t_{pd\pi}^2/\Delta_{pd}$.
The resulting bands near the Fermi energy have mainly Ru $d$ character:
% results are two {\color{red}\xcancel{characteristic}} types of bands:
a d$_{xy}$ band with dispersion that is isotropic in the $x,y$ plane, and a pair of bands, d$_{xz}$ and d$_{yz}$ with very anisotropic dispersion. The  d$_{xz}$ band disperses along the $x$ direction and is flat in the $y$ direction since the matrix element with the oxygen p states is zero, whereas the d$_{yz}$ band disperses in the $y$ direction and is flat in the $x$ direction.  At the $\Gamma$ point, d$_{xz}$ and d$_{yz}$ are degenerate with energy shifted up relative to the d$_{xy}$ band at $\Gamma$, since the hopping matrix element t$_{pd\pi}$ mixes the apical oxygen p states with the d$_{xz}$ and d$_{yz}$ bands but does not affect the d$_{xy}$ bands.

The multilayer systems consist of $n$ Ru-O layers
%which are strongly through the O atoms between the Ru-O layers,
as illustrated by the triple layer-system in Fig.~\ref{fig:SrRuO-structure}. Since each $n$-layer unit is weakly coupled to the adjacent $n$-layer unit, the final dispersion is essentially two-dimensional.
However, within $n$-layer each unit, the hopping through the apical oxygen is comparable to the hopping within the planes.  This leads to $n$ copies of the bands of a single layer which are split by the interlayer hopping into bonding and antibonding for two layers; bonding, non-bonding, and antibonding for three layers; and so forth.
The example of three layers is shown in the left side of Fig.~\ref{fig:tb-progression}, where the three d$_{xz}$ and  d$_{yz}$ states per Ru atom form bands with large splitting due to the hopping between the three Ru-O layers.  In contrast, the d$_{xy}$ states form three bands, which are degenerate in this model.

At the right of Fig.~\ref{fig:tb-progression} is the consequence of doubling of the unit cell in real space.  This is caused by the rotations of the O octahedra indicated by the arrows in  Fig.~\ref{fig:SrRuO-structure}. Since the bands are plotted in the smaller BZ, they appear complicated, and the relation to the unfolded bands is very useful for understanding.  Most relevant for our purposes is that the bands at the R point are folded to the $\Gamma$ point in the reduced BZ. In the original BZ (which applies if there are no rotations of the octahedra), the maxima in the bands are at the R point and there are no bands near the zone center $\Gamma$ in the upper energy range.  It is only because of the rotations that there are bands with maxima at the zone center $\Gamma$ that curve downward (hole-like).  An interesting consequence is that the intensity of the folded bands in a measurement is proportional to the amount of rotation and may appear to be weak in the actual experiments.

There are other effects that are not shown but can be understood qualitatively, e.g., magnetism.  In a tight binding model, this is simply a splitting of the majority and minority spin states leading to two sets of bands shifted almost rigidly, like those shown in  Figs.~\ref{fig:Exp_Geo_bands_tot} and \ref{fig:PBEsol-U2}. Since there is a large magnetization, the majority spin bands are almost filled and the Fermi energy must be near the top of the bands.
Thus, the majority spin bands are expected to be hole-like bands around $\Gamma$ with energies at or near the Fermi energy.  In contrast, the minority spin bands have reduced occupation with the Fermi energy in the lower parts of the bands.

\section{DFT Calculations and Computational details}
\label{sec:DFT- calcs}

\begin{figure}
\begin{center}
\includegraphics[width=0.45\textwidth]{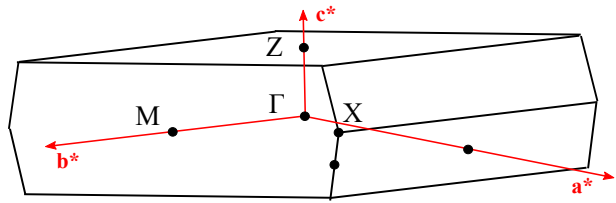}
\end{center}
\caption{BZ for Sr$_4$Ru$_3$O$_{10}$
structure as described in the text, where a*, b*, c* are the reciprocal lattice vectors for the base-centered orthorhombic structure with space group {\it Cmce} (64). Since the dispersion is almost two-dimensional, the points are labeled using the notation for a square lattice and the band structure plots are for lines that connect $\Gamma$ -- X -- M -- $\Gamma$ points in the $k_z = 0$ plane and $\Gamma$ -- Z on the $k_z$ axis.}
%The green lines are ones for plots sent by Garu in some previous calculations.
 \label{fig:BZ-high-symmetry-points-GG-RMM-both}
\end{figure}

There is much previous work on the bands of the ruthenates including Sr$_2$RuO$_{4}$  and Sr$_3$Ru$_2$O$_{7}$ which has a structure similar to Sr$_4$Ru$_3$O$_{10}$.
Density functional calculations have proven very successful in describing
the overall features of the observed bands~\cite{MMalvestuto_2011,CBergemann_2003,AMackenzie_2003,DJSingh_2001,MPAllan_2013,ATamai_2017}.
In all cases, the bands near the Fermi energy have mainly Ru $t_{2g}$ $d$ character with characteristic forms determined by only a few parameters, as described in Sec.~\ref{sec:Simplified-model}.  For the non-magnetic materials, the Fermi surfaces are determined by the band fillings and sensitive only to the relative energies of the $d_{xy}$ and the $d_{xz}$, $d_{xy}$ bands. There are effects of correlation that cause narrower bandwidths, however, the Fermi surfaces are not greatly affected because the Fermi energy is determined by band filling and the volume is constrained by the Luttinger theorem \cite{PhysRev.119.1153}. However, there is a qualitative difference for ferromagnetic  Sr$_4$Ru$_3$O$_{10}$.  The bands for the two spins have very different fillings and very different Fermi surfaces.  The results depend on the magnitude of the spin splitting which is sensitive to the functional.

We carried out first-principles calculations using the pseudopotential, plane-wave implementation of density functional theory in the Quantum ESPRESSO distribution~\cite{Giannozzi_2009, Giannozzi_2017}.
% to study the electronic structure and magnetism of the three-layered ruthenate Sr$_4$Ru$_3$O$_{10}$.
All calculations were performed using the PBEsol exchange-correlation functional~\cite{PhysRevLett.100.136406}  and ultrasoft pseudopotentials~\cite{PhysRevB.41.7892} with Sr(4s, 4p, 5s), Ru(4s, 4p, 5s, 4d), and O(2s, 2p) valence states~\cite{DALCORS_UPF, DALCORSO2014337}. The wavefunctions were expanded in plane waves with a cutoff of 75 Ry for the kinetic energy and 600 Ry for the charge density. Gaussian smearing with a broadening of 0.007 Ry was used in all our calculations. We have used the experimental lattice parameters and atomic positions taken from Crawford {\it et al}.~\cite{MKCrawford_2002}.  In that reference, the primitive cell contains two triple layers with 68 atoms per cell, but we have used the simplification in which the two triple layers are equivalent (which was suggested as a possibility~\cite{MKCrawford_2002}) in which case it is a base-centered orthorhombic structure with 34 atoms per cell with space group {\it Cmce} (64).  Given the proportions of a, b, and c of the unit cell (roughly thrice long along c than a and b), integrations over the BZ of this lattice were determined to be 8 x 8 x 3 Monkhorst-Pack~\cite{PhysRevB.13.5188} k-point meshes.  The grids were 32 x 32 x 8 for the Fermi surface calculations.
Structural relaxations were performed with atomic forces converged to within 1.0 x 10$^{-5}$ eV/Å, while energies were converged to within 1.0 x 10$^{-11}$ eV.
The results presented here are for the optimized structure with the above criteria.

In all cases, the lowest energy is a ferromagnet with large spin polarization in general agreement with the experiment.  The moment calculated for the optimized structure is $1.77 \mu_B/\text{Ru}$, which is in agreement with the experimental values in the range $\sim 1.0$ to $\sim 1.7\mu_B/\text{Ru}$~\cite{MKCrawford_2002,GCao_2003,ZQMao_2006,MZhou_2005,FWeickert_2017, FForte_2019}. The moments are largest on the central layer in agreement with a spin-polarized neutron diffraction study~\cite{FForte_2019}. Since there is large spin polarization, there is a large difference in occupation of the majority and minority spin bands and we expect qualitative effects on the states at the Fermi energy.

The BZ for the actual orthorhombic crystal structure is shown in  Fig.~\ref{fig:BZ-high-symmetry-points-GG-RMM-both}.  The calculated bands are plotted for the lines $\Gamma$--X--M--$\Gamma$--Z connecting the points shown in the figure.  However, since there is only small dispersion in the direction perpendicular to the layers, the bands are almost two dimensional and we have used the notation for the points in the BZ zone appropriate for a two dimensional square lattice.  This is the notation used for the ARPES experiments~\cite{PNgabonziza_2020} where the data are analyzed in terms of a two-dimensional square BZ. We have investigated the kz dispersion in other parts of the BZ and we find that the dispersion at all places studied is less than that along the $\Gamma$ to Z line which is shown in the following figures for the band structures. 
\subsection{Non-spin-polarized calculations}

\begin{figure}
\begin{center}
\includegraphics*[width=0.5\textwidth]{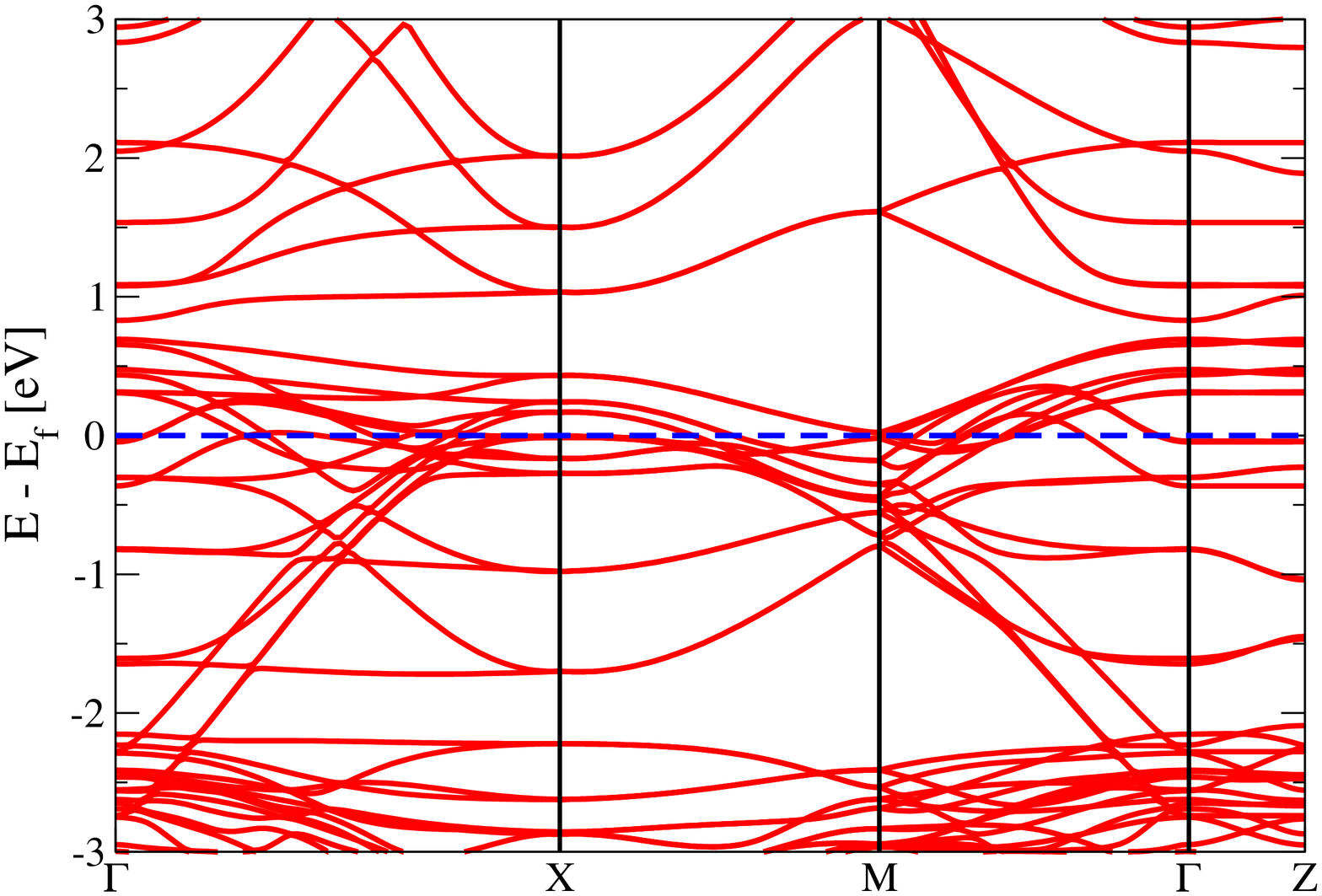}
\includegraphics*[width=0.5\textwidth]{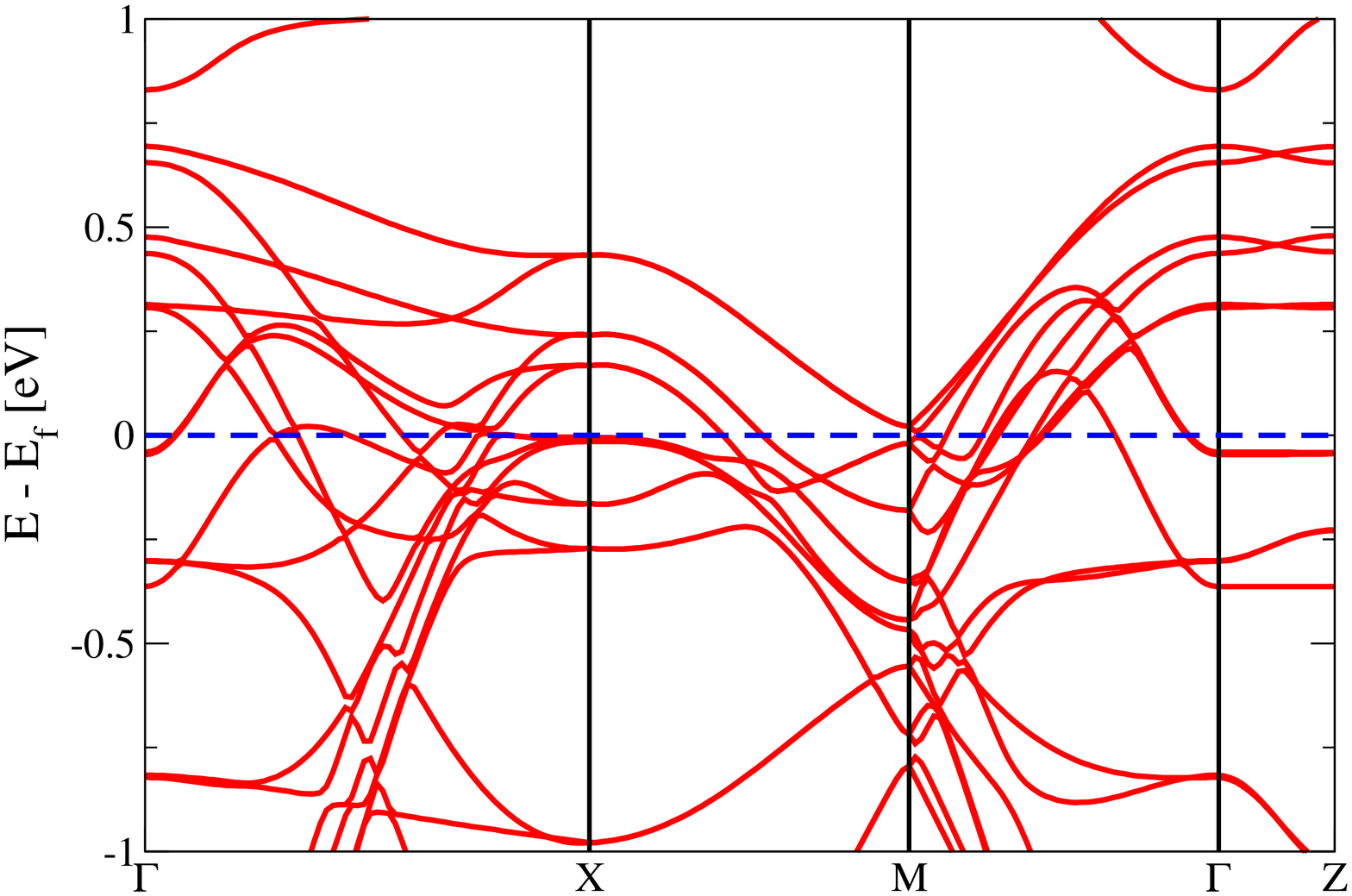}
\end{center}
\caption{Non-spin-polarized bands of Sr$_4$Ru$_3$O$_{10}$ plotted along the lines connecting the points shown in Fig.~\ref{fig:BZ-high-symmetry-points-GG-RMM-both}.
% (I got Garu's results confused. He says it is the same as Sept. 21, and very similar or identical to ones sent on March 9.)
The Fermi energy is set to zero and indicated by the dashed blue line.
The bands at the bottom of the top figure have mainly oxygen in character;
those from ~-2 eV to ~1 eV are Ru t$_{2g}$ d bands mixed with O p states.  The steeply increasing bands are due to the d$_{xy}$ states, whereas the d$_{xz}$ and d$_{yz}$ form pairs of states degenerate at $\Gamma$ with one almost flat and the other with large dispersion along the $\Gamma$ - X direction. These states in different layers are strongly coupled leading to bonding, non-bonding and antibonding bands at approximately -1.6, -0.8 and 0.3 eV at $\Gamma$.}
%mixed with other bands around the Fermi energy.}
\label{fig:bands-NM}
\end{figure}

 We first present the calculated band structures and Fermi surfaces of
 %the three-layered
 Sr$_4$Ru$_3$O$_{10}$ for the case where the system is constrained to have no spin polarization, %using PBEsol functional
 as shown in Fig.~\ref{fig:bands-NM}  and the left side of Fig.~\ref{fig:Fermi_surface_up-down} respectively. The conclusion in Sec.~\ref{sec:Exp-comparison} is that the results do not agree with the experiment~\cite{PNgabonziza_2020}.  Nevertheless, this is important to show the large difference from the spin-polarized results and it provides a simpler picture of the bands that are a guide to understanding the more complicated case where there is no spin-degeneracy and there are twice as many bands.

The position of the bands relative to the Fermi energy is determined by the fact that there are four electrons in the 6 $t_{2g}$ states per Ru atom, {\it i.e.}, 2/3 filling of the $t_{2g}$ bands.  As we can see from the upper part of Fig.~\ref{fig:bands-NM}, the
	bands below around -2 eV are primarily oxygen in character and the bands above approximately 1 eV are $e_g$ Ru~$d$ states.  The main focuses here are the $t_{2g}$ bands in the range -2 to 1 eV and the bands that extend to below -2 eV in the range of the oxygen bands.  The $t_{2g}$  bands are mainly Ru $d$ character and they look remarkably like the tight-binding bands shown in Fig.~\ref{fig:tb-progression}.
	
This is shown most clearly by the flat bands along $\Gamma$ to X that identify the bonding, non-bonding and antibonding
d$_{xz}$ and d$_{yz}$ bands and the images that are folded into the small BZ very much like that shown in the right side of Fig.~\ref{fig:tb-progression}. The magnitude of the splitting of the bonding and antibonding bands is approximately 1.3 eV which is very similar to the value of approximately 1.2 eV calculated for Sr$_3$Ru$_2$O$_7$~\cite{DJSingh_2001}.
The bonding and non-bonding $d_{xz}$ and $d_{yz}$ bands are evident as shown by the almost flat bands; the antibonding bands are not as clear since they are mixed with other bands that are folded into the BZ. The bonding and non-bonding combination of the d$_{xz}$ and d$_{yz}$ states in the three layers are mainly full and the Fermi energy is in the range of the anti-bonding bands, where there are many bands.

 The bands that increase steeply upward going away from $\Gamma$ both toward X and toward M have $d_{xy}$ character.  There are three bands that are almost decoupled from one another with one band for the central layer and two forming a degenerate pair at a different energy since the potentials at the central layer and outer layers are different.

The calculated Fermi surfaces shown in the left side of Fig.~\ref{fig:Fermi_surface_up-down} are qualitatively similar to those for
Sr$_3$Ru$_2$O$_7$~\cite{ATamai_2008}; the shapes of the contours are similar, but the number of bands and the sizes of the pieces of Fermi surfaces are different for three layers compared to two in Sr$_3$Ru$_2$O$_7$.   The flat pieces of the Fermi surface are due to the $d_{xz}$ and $d_{yz}$ bands, which are almost flat as a function of $k_x$ for  $d_{yz}$ and $k_y$ for  $d_{xz}$.  These flat pieces of the Fermi surface show how the lattice is oriented since they are perpendicular to the rows and columns of Ru atoms.  Along the boundaries of the BZ, the Fermi surfaces are complicated due to many bands crossing the Fermi energy, as can be seen in Fig.~\ref{fig:bands-NM}.  We have not tried to interpret these since they depend upon the details and are not relevant to the spin-polarized ferromagnetic system considered next.

\subsection{Spin-polarized calculations}

\begin{figure}[htbp!]
\begin{center}
\includegraphics*[width=0.5\textwidth]{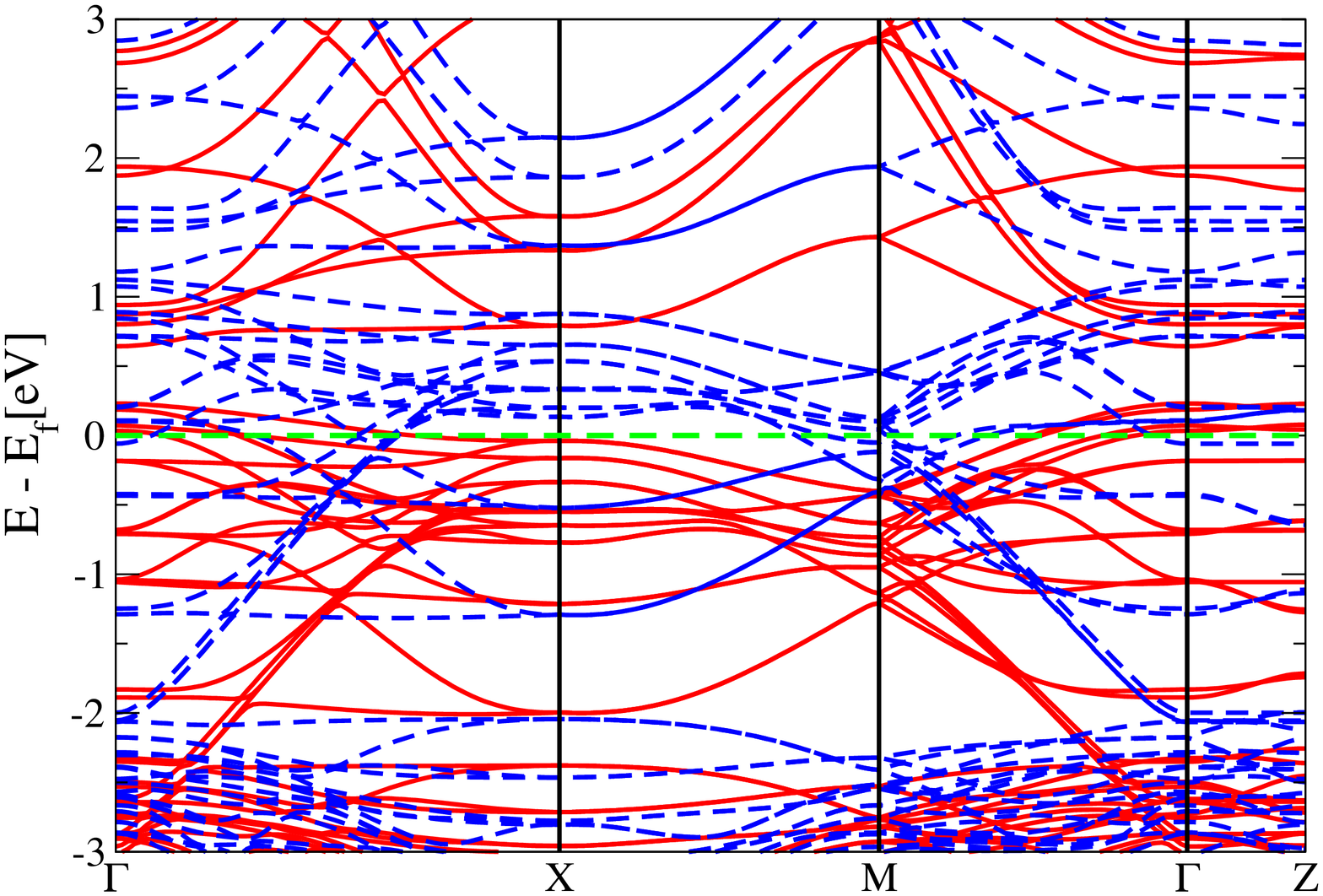}
\includegraphics*[width=0.5\textwidth]{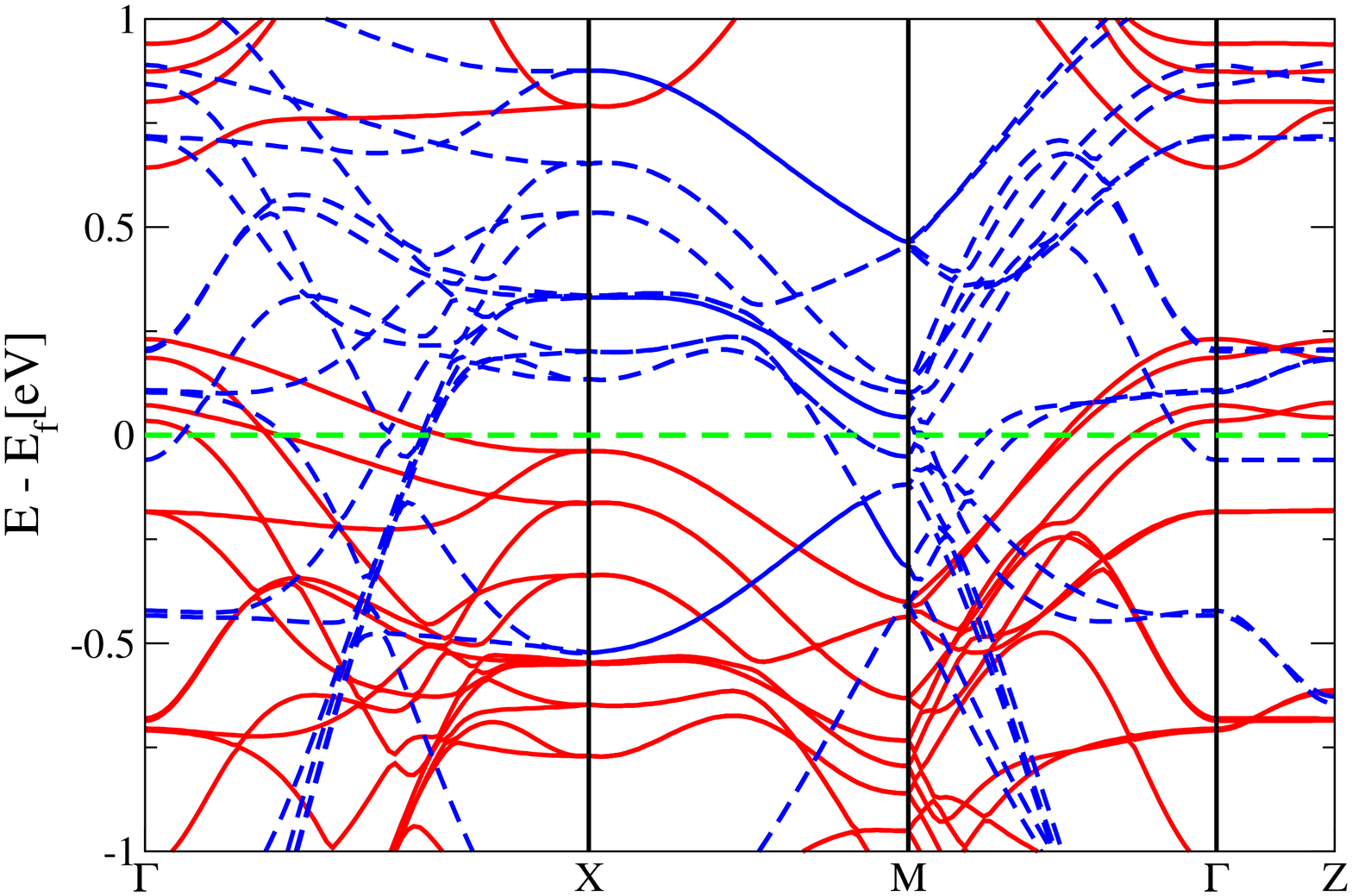}
\end{center}
\caption{Spin polarized bands of Sr$_4$Ru$_3$O$_{10}$  plotted along the lines connecting the points shown in Fig.~\ref{fig:BZ-high-symmetry-points-GG-RMM-both}. The Fermi energy is set to zero and indicated by the dashed green line.
  There are two sets of bands for the majority (red solid lines) and minority bands (blue dashed lines) spin. The bands from the two spins are shifted almost rigidly, except that lower energy parts of the majority spin bands are strongly mixed with oxygen bands.
  %, and effects of interactions between the many bands.
  As discussed in the text, the bands for each spin are very similar to the bands for the non-spin-polarized case in Fig.~\ref{fig:bands-NM}.
  The Fermi surfaces for the two spin states are shown in the center and right figures in Fig.~\ref{fig:Fermi_surface_up-down}.}
\label{fig:Exp_Geo_bands_tot}
\end{figure}

\begin{figure*}
\begin{center}
\includegraphics[width=0.32\textwidth]{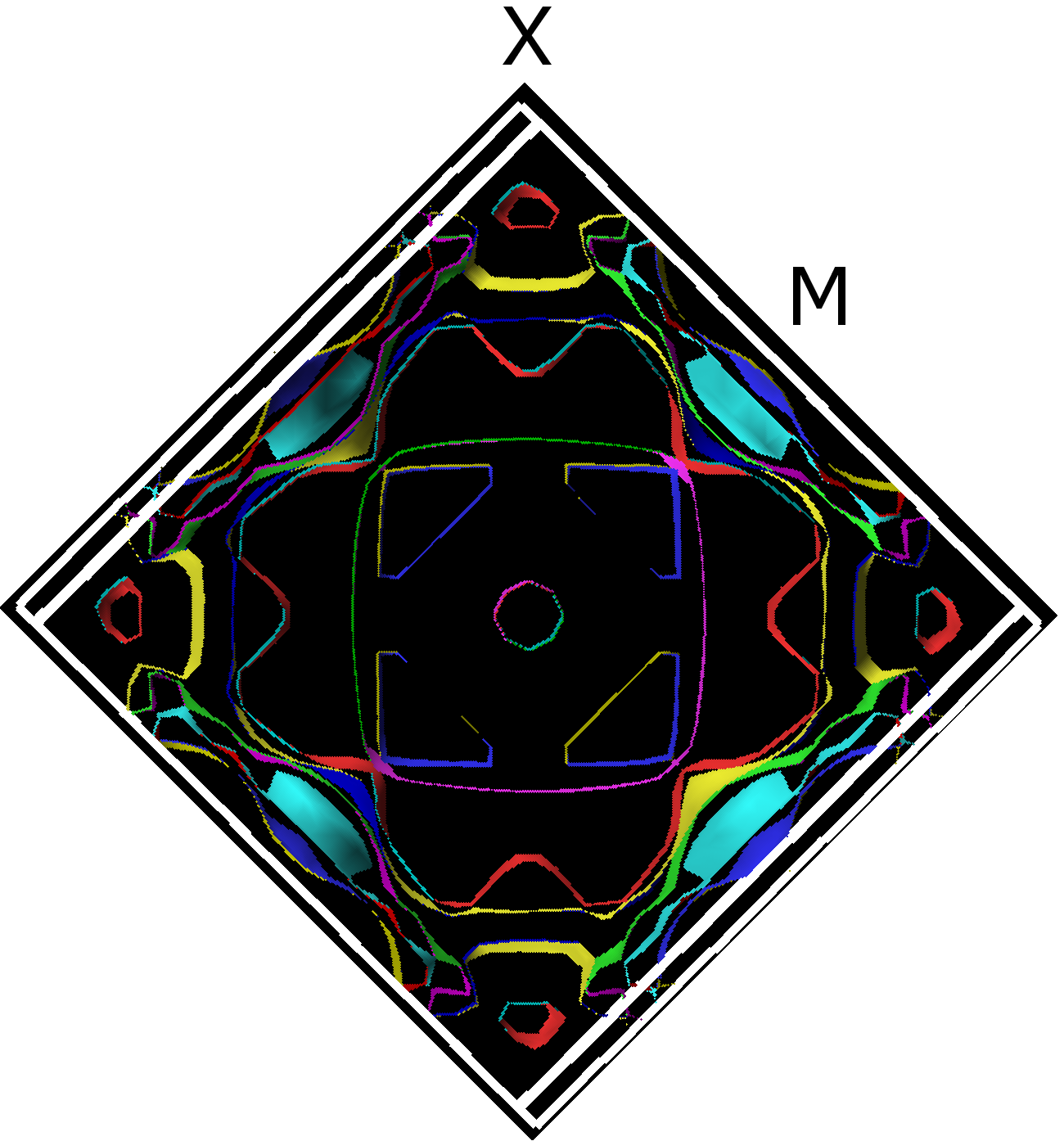}%
\includegraphics[width=0.32\textwidth]{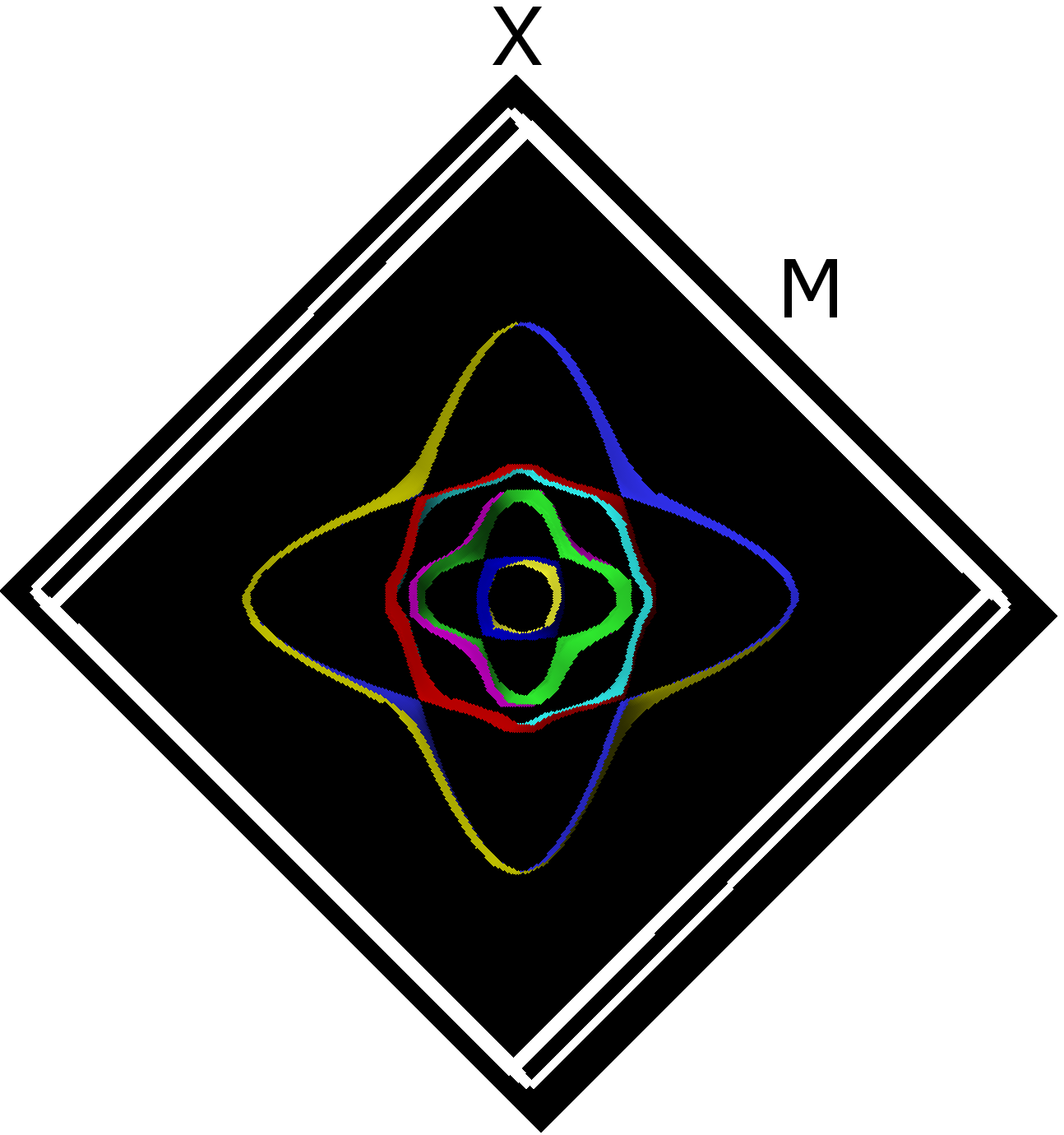}%
\includegraphics[width=0.32\textwidth]{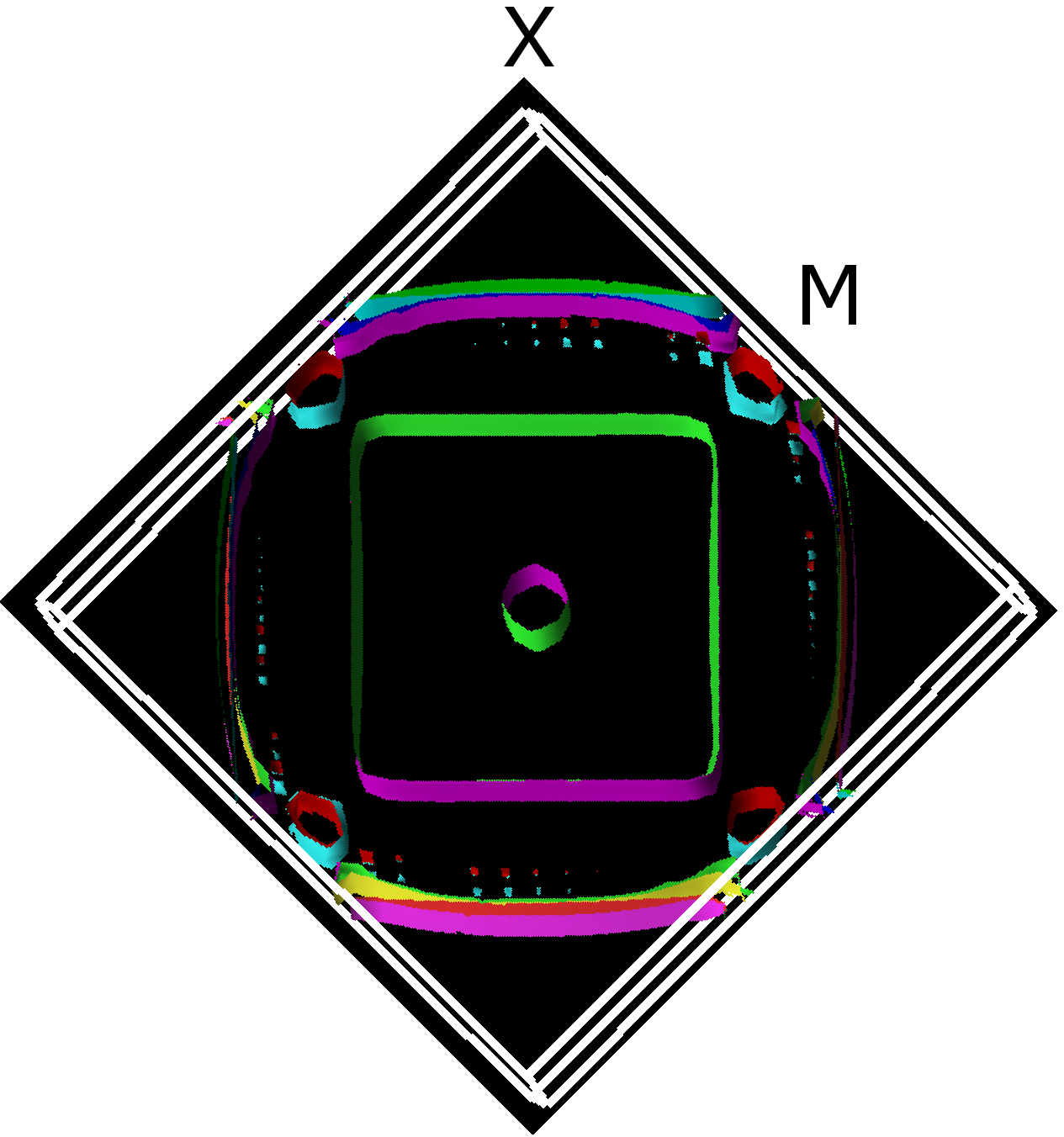}%
\end{center}
\caption{Fermi surfaces for Sr$_4$Ru$_3$O$_{10}$ resulting from two different calculations.  At the left is the result for a calculation constrained to have no spin polarization corresponding to the bands in Fig.~\ref{fig:bands-NM}.
The two figures in the center and right are the Fermi surfaces for the majority and minority spins in the spin-polarized calculation corresponding to the bands in Fig.~\ref{fig:Exp_Geo_bands_tot}. As described in the text, the spin-polarized calculations explain the major features of the ARPES results shown in Fig.~\ref{fig:fs-ARPES}.}
 \label{fig:Fermi_surface_up-down}
\end{figure*}

The calculated band structures and Fermi surfaces for the same crystal structure using the same PBEsol functional, but allowing for spin polarization are shown in Fig.~\ref{fig:Exp_Geo_bands_tot}  and  Fig.~\ref{fig:Fermi_surface_up-down} respectively.

The ground state is found to be ferromagnetic so that there are two sets of bands for the two spins.  The bands have very nearly the same form as in the non-spin-polarized calculation shown in Fig.~\ref{fig:bands-NM},
except shifted relative to one another.  As expected from the large polarization, the shift of the bands is large which changes the picture qualitatively from a nonmagnetic state. The majority spin bands are shifted down and are almost filled so that they have hole-like character near the Fermi energy.  These are the dominant bands near the zone center and there are several bands near the Fermi energy that are quite flat, especially along the line from $\Gamma$ to X.

In contrast, the minority spin bands shifted up so that they are approximately $1/3$ filled and have electron-like character dispersing upward across the Fermi energy as shown in Fig.~\ref{fig:Exp_Geo_bands_tot}. There are three $d_{xy}$ bands for the three layers that disperse steeply upward across the Fermi energy: two on the outer layers and very nearly degenerate, and one on the central layer with slightly different energy. These give rise to the nearly circular pieces of the Fermi surface shown in the right side of Fig.~\ref{fig:Fermi_surface_up-down}.
The non-bonding combination of $d_{xz}$ and $d_{yz}$ are the pair of bands that are degenerate at $~-0.4$ eV, and which disperse upward to form the square shaped piece of Fermi surface.  The character of the minority spin bands is essentially the same for all the spin-polarized calculations reported here and shown in  Figs.~\ref{fig:Exp_Geo_bands_tot} and \ref{fig:PBEsol-U2}.

%\begin{widetext}

%\end{widetext}

The Fermi surfaces for the majority and minority spins are shown in the center and right figures in Fig.~\ref{fig:Fermi_surface_up-down}. The surfaces are qualitatively different from those for the non-spin-polarized system shown at the left in  Fig.~\ref{fig:Fermi_surface_up-down}. The surfaces for each spin state are much simpler and have more straightforward interpretations. Because of the large polarization, pieces of the Fermi surface near the zone center are mainly due to the majority spin bands, whereas pieces nearer than the BZ boundary are mainly due to the minority spins.

The pieces of the Fermi surface shaped like extended ovals with the long direction pointing toward the X points at the BZ corners are due to the majority spin $d_{xz}$ and $d_{yz}$ bands, and they are very similar to $t_{2g}$ bands found in many two dimensional systems, such as oxide interfaces in~\cite{ZZhong-PhysRevB.87.161102}. However, in the present case, the position of the majority spin bands relative to the Fermi energy is very sensitive to the functional used in the calculation. This is shown in the calculations described below where the Hubbard $U$ correction leads to a reduction of the size of the ovals until they disappear completely. Comparison of the bands with ARPES experiments is presented in Sec.~\ref{sec:Exp-comparison}.
\CBK 
%Therefore, we do not try to analyse the detailed features of the majority spin bands.

\subsection{Spin-orbit coupling}

\begin{figure}[!b]
\begin{center}
\includegraphics[width=0.5\textwidth]{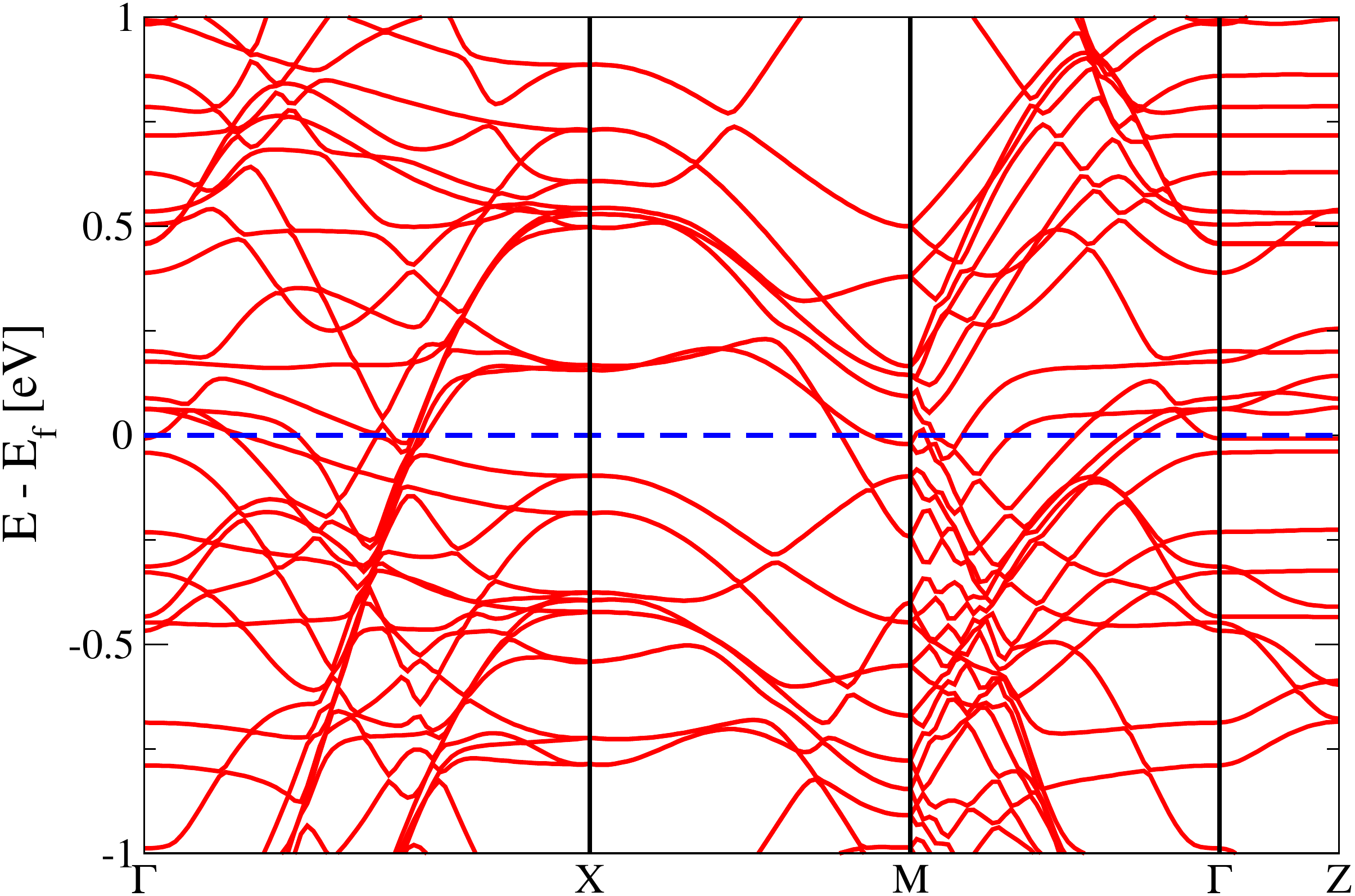}
\end{center}
\caption{Bands of Sr$_4$Ru$_3$O$_{10}$ with the inclusion of spin-orbit coupling plotted along the lines connecting the points shown in Fig.~\ref{fig:BZ-high-symmetry-points-GG-RMM-both}. The Fermi energy is set to zero and indicated by the dashed blue line. The changes are mainly small shifts of bands and small gaps; however, this has large effect on the narrow bands near the Fermi energy, which can be seen in the Fermi surfaces in Fig.~\ref{fig:Fermi_surface_SOC}.}
	%Spin Polarized bands of Sr$_4$Ru$_3$O$_{10}$  the same %as the bands in Fig.~\ref{fig:Exp_Geo_bands_tot} but %including spin-orbit coupling. The changes are mainly %small shifts of bands and small gaps. The Fermi %surfaces are shown in %Figs.~\ref{fig:Fermi_surface_up-down} and %\ref{fig:Extended-data-Prosper-thesis}, which can be %recognized as  close to a sum of the Fermi surfaces for %majority and minority spins, but with significant %differences due to the small changes near the Fermi %energy.}
\label{fig:Exp_Geo_bands_SOC-1to1}
\end{figure}

\begin{figure}[!b]
\begin{center}
\includegraphics[width=0.32\textwidth]{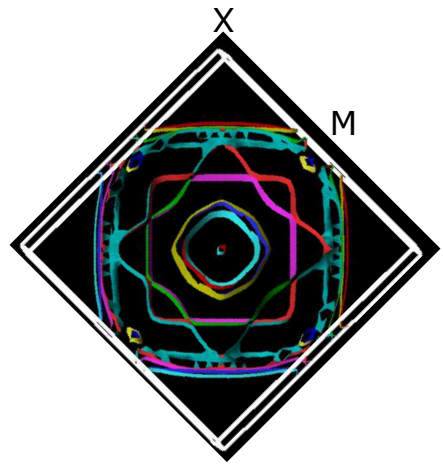}
\end{center}
\caption{Fermi surfaces for Sr$_4$Ru$_3$O$_{10}$ including spin-orbit coupling corresponding to the bands in Fig.~\ref{fig:Exp_Geo_bands_SOC-1to1}; the surfaces are close to a sum of those for the majority and minority spins shown respectively in the center and right figures in Fig.~\ref{fig:Fermi_surface_up-down}, with changes mainly at the places where there are narrow bands.\CBK}
 \label{fig:Fermi_surface_SOC}
\end{figure}

The calculated band structures and Fermi surfaces for the same crystal structure using the same PBEsol functional with the inclusion of the spin-orbit couplings are shown in
Fig.~\ref{fig:Exp_Geo_bands_SOC-1to1} and Fig.~\ref{fig:Fermi_surface_SOC} respectively. Since the spin-orbit effects are small compared to the bandwidths, the bands are still primarily majority and minority spin.  By comparing the bands in Figs.~\ref{fig:bands-NM} and \ref{fig:Exp_Geo_bands_SOC-1to1}, we can see that the effects are mainly to open small gaps and remove band crossings.  The effects are most significant in cases where there are band crossings for narrow bands near the Fermi energy, which is illustrated by the Fermi surfaces shown in  Fig.~\ref{fig:Fermi_surface_SOC}.  Note that in all the calculations, there are large square parts of the Fermi surfaces, which is the most prominent feature found in the ARPES experiments as described in Sec.~\ref{sec:Exp-comparison}.
\CBK

\subsection{Including a Hubbard $U$ correction}

We also performed DFT + $U$ calculation by applying a Hubbard $U$ correction on the Ru 4d states~\cite{PhysRevB.44.943}.
%based on DFPT \cite{PhysRevB.98.085127}.
Calculations with similar functionals with and without an added $U$ have been reported for  Sr$_2$RuO$_{4}$~\cite{HLHuang_2020, Mario-1999}, Sr$_3$Ru$_2$O$_{7}$~\cite{David-2020}, and SrRuO$_3$~\cite{Igor-2014}.  The  values of $U$ used in the previous studies~\cite{HLHuang_2020, Igor-2014, Mario-1999, David-2020, Cuoco-PhysRevB.57.11989, Liebsch-PhysRevLett.84.1591, Pchelkina-PhysRevB.75.035122, Malvestuto-PhysRevB.83.165121} were in the region 1.5–4.0 eV. Here we present results for two moderate values, $U$ = 1.0 eV and $U$ = 2.0 eV, which show the large effects on the electronic properties of Sr$_4$Ru$_3$O$_{10}$.

\begin{figure}
	\begin{center}
		\includegraphics*[width=0.5\textwidth]{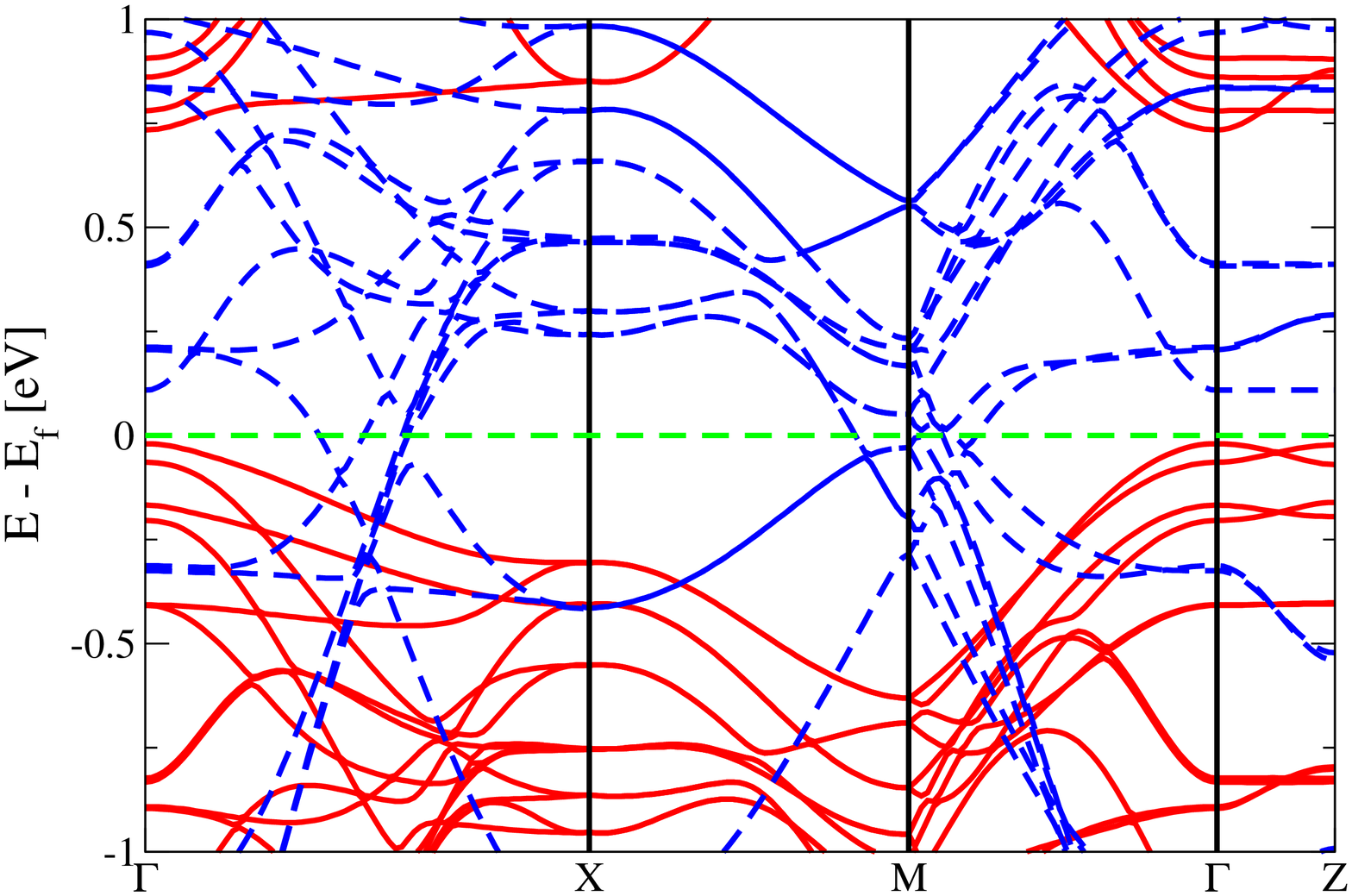}
		\includegraphics*[width=0.5\textwidth]{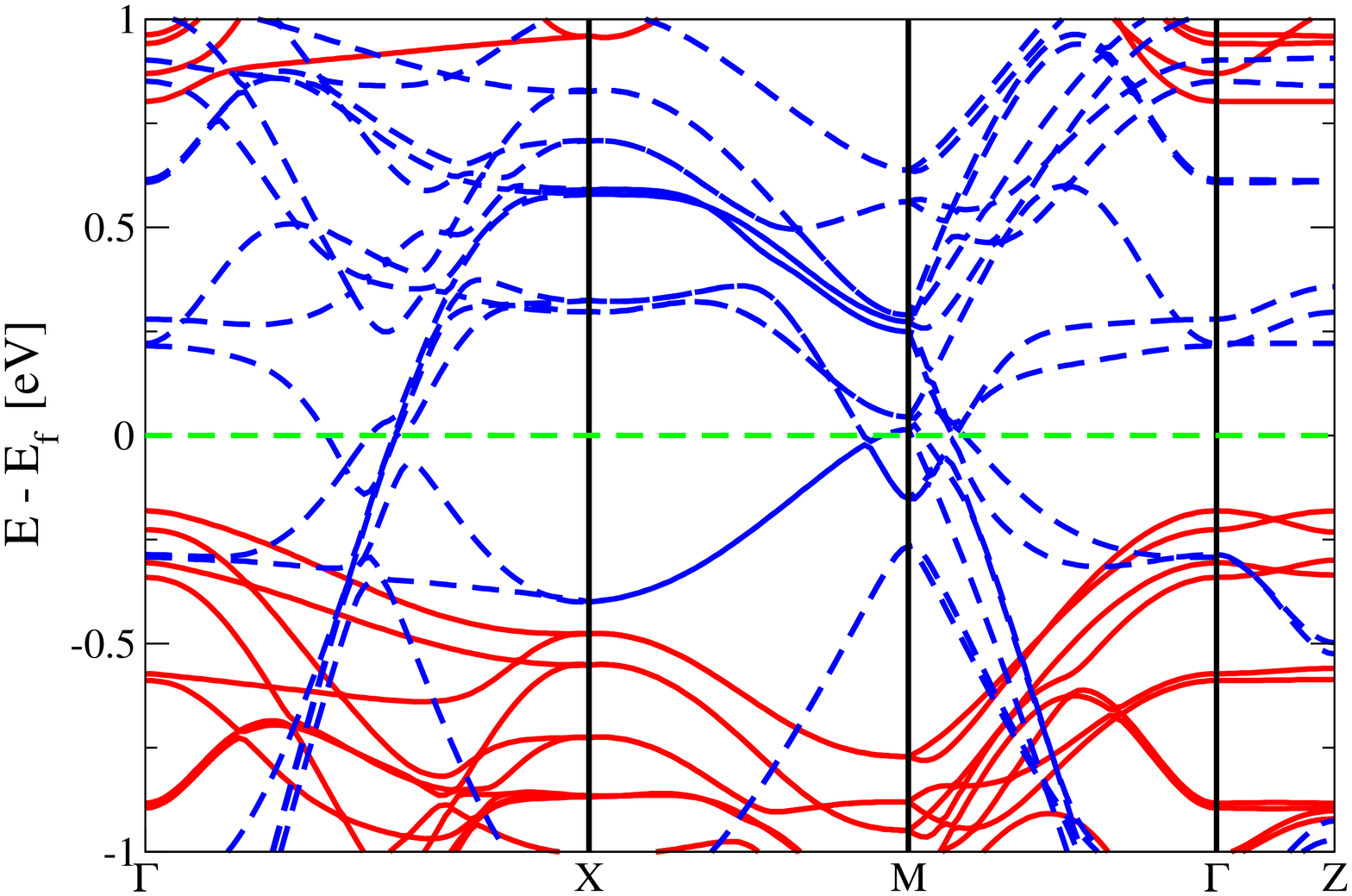}
	\end{center}
	\caption{Spin polarized bands of Sr$_4$Ru$_3$O$_{10}$ with the inclusion of the Hubbard $U$ interaction on the Ru atoms plotted along the lines connecting the points shown in Fig.~\ref{fig:BZ-high-symmetry-points-GG-RMM-both}. The Fermi energy is set to zero and indicated by the dashed green line. The upper figure is for $U=1.0$ eV and the lower figure is for  $U=2.0$ eV. As $U$ increases the spin splitting increases and at $U=1.0$ eV the majority spin band is filled and just touches the Fermi energy. For larger $U$ the system is fully polarized - a half-metallic system with only minority spin states at the Fermi energy. The Fermi surfaces for the minority spin for $U=2.0$ eV are shown in Fig.~\ref{fig:fs-U2}; the surfaces for $U=1.0$ eV are very similar and are not shown. }
	\label{fig:PBEsol-U2}
\end{figure}

The calculated band structures using the PBEsol + $U$ functional with the inclusion of Hubbard $U= 1.0$ eV and $2.0$ eV are shown in Fig.~\ref{fig:PBEsol-U2}. The crystal structure is the same as in the calculations for $U =0$ eV. \CBK As expected, the addition of a repulsive interaction $U$ for electrons in the Ru $3d$ orbitals leads to a larger splitting of the bands.
For $U=1.0$ eV  the highest states of the majority spin band just touch the Fermi energy.  This is the dividing point where the nature of the Fermi surface changes. For $U< \approx 1.0$ eV, there are pieces of Fermi surface which are hole-like around the zone center for the majority-spin bands. For $U> \approx 1.0$ eV the majority spin bands are completely filled and there are only minority spin bands at the Fermi energy.  As $U$ increases above this value, there are small quantitative changes in the shapes of the bands
%, but there are only small changes in
and the Fermi surfaces since the filling of the minority spin bands is fixed at 1/3.  This is illustrated in Fig.~\ref{fig:PBEsol-U2} where we can see that the majority spin bands shift downward relative to the Fermi energy for $U=2.0$ eV, but the minority spin bands stay almost the same.

\begin{figure}
	\begin{center}
		\includegraphics[width=0.32\textwidth]{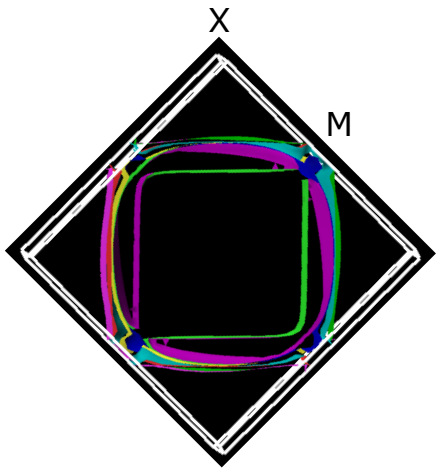}
	\end{center}
	\caption{The Fermi surface for the minority spin bands calculated with the  inclusion of Hubbard $U$. The results shown are for $U=2.0$ eV and the Fermi surface is essentially the same for $U=1.0$ eV.  There are no bands for the majority spin that cross the Fermi energy, but there are bands near the $\Gamma$ point that are close to the Fermi energy as shown in Fig.~\ref{fig:PBEsol-U2}.}
	\label{fig:fs-U2}
\end{figure}

The minority spin Fermi surface for $U=2.0$ eV is shown in Fig.~\ref{fig:fs-U2}.
It is essentially the same for $U=1.0$ eV (not shown) and, in fact, the minority spin Fermi surface is qualitatively the same for $U=0$, which can be seen by comparing Fig.~\ref{fig:fs-U2} with the Fermi surface at the right side in Fig.~\ref{fig:Fermi_surface_up-down}.  Thus a definitive result of our calculations is that for any reasonable functional, the result for the minority spin Fermi surface has qualitatively the same form.  As discussed in the following section this is the major feature seen in the experimental ARPES data.

\section{Comparison with ARPES experiments}
\label{sec:Exp-comparison}

In this section we compare the theoretical calculations presented above with recent ARPES experiments~\cite{PNgabonziza_2020} which are focused on the Fermi surface and the bands near the Fermi energy.
Since the experiments do not resolve the spin, there is no direct evidence of the ferromagnetic moments nor of the difference between majority and minority spins. Thus the conclusions regarding the effects of spin polarization are
%in terms of the predictions of measured Fermi surfaces and bands including both spins.
prediction of the theory.  Nevertheless the experimental results~\cite{PNgabonziza_2020} provide definitive tests of the theory because the energies of the bands are greatly modified by the polarization, and we can compare the predicted and measured energies of the bands and shapes of the Fermi surfaces.\CBK

In the comparison with the ARPES experiments, it is important to keep in mind that the experiments may be affected by changes of the structure at the surface, and there may be surface states like those found, for example, in ARPES for Sr$_2$RuO$_4$~\cite{ADamascelli_2000}. However, careful studies to distinguish bulk and surface states are not reported in the experimental paper~\cite{PNgabonziza_2020}.

\subsection{Fermi surfaces}

\begin{figure}
	\begin{center}
		\includegraphics[width=0.50\textwidth]{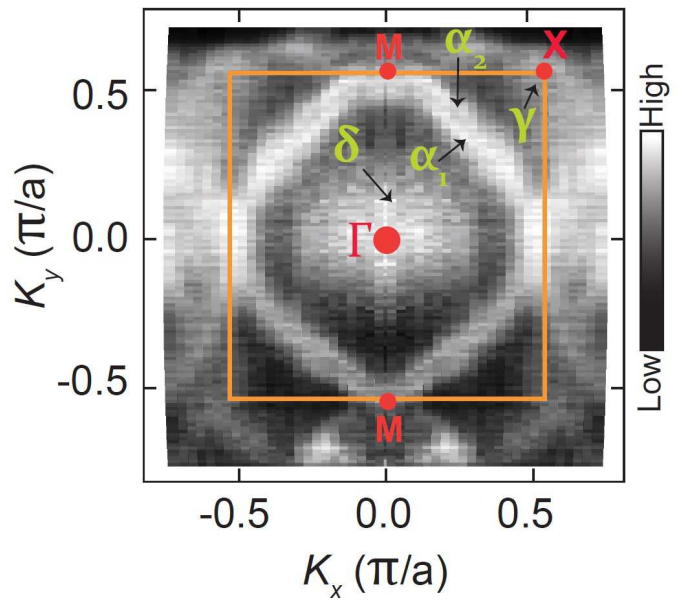}
	\end{center}
	\caption{The experimental Fermi surface reported in \cite{PNgabonziza_2020}.  This is taken from Fig. 1a of \cite{PNgabonziza_2020} with no changes.  As explained in the text, it must be turned by 45 degrees to compare with the present theoretical calculations.  The white areas correspond to the large intensity at the Fermi energy, and it is immediately apparent that the primary feature - the large roughly square shape - agrees well
		with the calculated Fermi surface for the ferromagnetic calculations (Fig.~\ref{fig:Fermi_surface_SOC} or Fig.~\ref{fig:fs-U2}) and it is qualitatively different from the calculated Fermi surface if the system is forced to have no spin polarization (the left side of Fig.~\ref{fig:Fermi_surface_up-down}).}
	\label{fig:fs-ARPES}
\end{figure}

First, we consider the Fermi surfaces because this is the information that shows most clearly the primary result of the present work: the large effect of the ferromagnetism on the electronic spectrum.  In Fig.~\ref{fig:fs-ARPES} is shown the intensity at the Fermi energy as a function of the momentum in two dimensions parallel to the layers.  The figure is a reproduction of Fig.~1a in \cite{PNgabonziza_2020}, where the data are presented with the square BZ oriented as shown (with the sides vertical and horizontal) which corresponds to the way the data was acquired. In order to compare with the present calculations which are presented in the conventional orientation of the BZ, the image must be turned by 45 degrees; this is only a choice of presentation that does not affect the conclusions~\footnote{In the experimental paper\cite{PNgabonziza_2020} it was stated that at their measurement temperature the structure was modified so that the BZ was turned by 45 degrees; however, this is not needed and the analysis of the experiments are consistent with the structure we use here.}.
The bright areas indicate states at or very near the Fermi energy.

The dominant feature in Fig.~\ref{fig:fs-ARPES} is the square shaped region indicated as $\alpha_1$ and $\alpha_2$.   This is in good agreement (when turned by 45 degrees) with the sets of bands in the Fermi surfaces for the minority spin calculated for the system with ferromagnetic order: the right side of Fig.~\ref{fig:Fermi_surface_up-down} for the calculation with U=0 or Fig.~\ref{fig:Fermi_surface_SOC} where the spin-orbit coupling is included, or Fig.~\ref{fig:fs-U2}, which shows the Fermi surface calculated with $U=2.0$ eV, which is essentially the same for all $U>1.0$ eV where there is only the minority spin Fermi surface.  As shown most clearly in Fig.~\ref{fig:fs-U2}, the calculations find this to be a combination of the nearly circular pieces of Fermi surface shown due to the three nearly degenerate $d_{xy}$ bands in the three Ru-O layers and the square shaped piece of Fermi surface due to the non-bonding combination of $d_{xz}$ and $d_{yz}$ states in a triple layer.
The character of the bands called $\alpha_1$ and $\alpha_2$ is supported by the other results in the ARPES experiments, which show that these are wide bands with a large slope in agreement with the calculated minority spin bands.
These are robust results found in all the spin-polarized calculations and clear support for the interpretation in terms of spin-polarized bands.

The observation that these bands have strong intensity in the measurements is additional evidence for the theoretical conclusion that they are due to minority spin bands.  This can be understood following the description of the bands in the pedagogical model in Sec.~\ref{sec:Simplified-model}, where it is clear that in the lower energy range the bands are properties of the lattice of Ru-O octahedra independent of whether the octahedra are rotated or not. This is the energy range applicable for the minority spin bands in the DFT calculations, which have reduced occupation (approximately 1/3) so that the Fermi energy is in the lower part of the minority-spin $t_{2g}$ bands.  In contrast the upper parts of the bands are ``folded'' into the smaller BZ as indicated in right hand side of Fig.~\ref{fig:tb-progression}.  This is the energy range relevant for the majority spin bands which are fully or nearly fully occupied in the DFT calculations.  Thus the intensity is non-zero only because of the rotations of the octahedra, and the intensity for these bands may be weak in the experimental measurements for momenta in the first BZ.
\CBK

Thus the major features in the ARPES data is well explained by the theoretical calculations with spin-polarization and it is quantitatively different from the Fermi surfaces predicted by the non-spin-polarised calculations shown in the left side of Fig.~\ref{fig:Fermi_surface_up-down}.  In that case, the occupation of the bands is 2/3, the Fermi energy is in the upper parts of the bands, and there are a host of pieces of Fermi surface (left side of Fig.~\ref{fig:Fermi_surface_up-down}) near the zone boundaries that are not seen in the experimental data.  They are analogous to results of the non-spin-polarized calculations and the experimental results for Sr$_3$Ru$_2$O$_7$~\cite{ATamai_2008}.

%The ARPES data is quantitatively different from the Fermi surface predicted in the system is constrained to have no spin polarization as shown in the left side of Fig.~\ref{fig:Fermi_surface_up-down}.  In that case the occupation of the bands is 2/3, the Fermi energy is in the upper parts of the bands, and there are a host of pieces of Fermi surface (left side of Fig.~\ref{fig:Fermi_surface_up-down}) near the zone boundaries that are not seen in the experimental data.  (They are analogous to results of the non-spin-polarized calculations and the experimental results forSr$_3$Ru$_2$O$_7$~\cite{ATamai_2008}.)

\subsection{Narrow bands near the Fermi energy}

The other prominent feature in Fig.~\ref{fig:fs-ARPES} is a broad area of intensity in the region around the zone center labeled $\delta$.  Even though there is no direct evidence in this figure for a Fermi surface, this indicates narrow bands at or near the Fermi energy within the energy resolution over a large area of the BZ around the zone center. This also is in qualitative agreement with the present calculations for the ferromagnetic system where the majority spin bands are filled or almost filled with the maxima at the zone center.  The theory predicts that these bands near the top of the majority spin bands should have weaker intensity in the experiment because they have non-zero intensity near the center of the BZ only because of the rotations of the octahedra.  The observations support this interpretation and are very different from the predictions of the non-spin-polarized calculation shown in Fig.~\ref{fig:bands-NM} where there are no hole-like bands around the zone center at the Fermi energy.
%
%\CRD DID NOT WORK ON THE REST OF THIS SECTION.  GARU IS WORKING ON IT.  IDEA: OTHER PAPERS (TAMAI?) SAY S-O IS THE SAME OR INCREASED AT THE SAME TIME AS BANDS ARE NARROWED BY COULOMB INTERACTIONS. WOULD HAVE BIG EFFECTS.\CBK

\begin{figure}
	\begin{center}
		\includegraphics[width=0.50\textwidth]{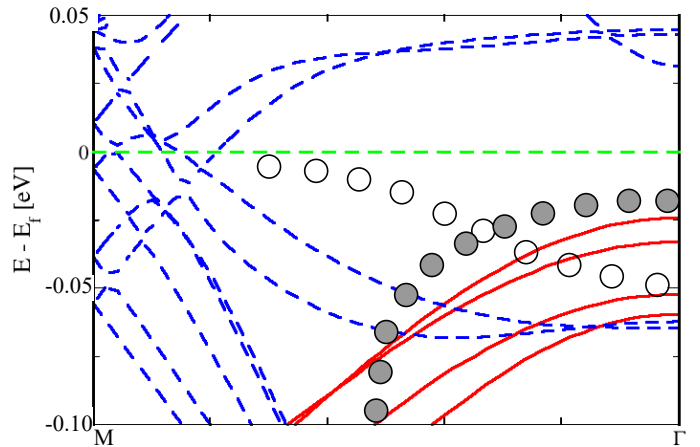}	
	\end{center}
	\caption{
%The experimental data are reported in \cite{PNgabonziza_2020}.  The middle top  and right top are taken from supplementary material Fig. S2 and Fig. 4c of \cite{PNgabonziza_2020} with no changes.  The black line on the middle top and black dotted line in the right top are hand drawn lines. The two experimental data are measured with a Photon energy of 60 eV in Linear Horizontal Polarization (LHP) and in Linear Vertical Polarization (LVP) respectively (see~\cite{PNgabonziza_2020} for details). The bottom figures are the
%\CRD Comparison of the calculated spin polarized bands of Sr$_4$Ru$_3$O$_{10}$ along the lines connecting $M$ and $\Gamma$ points with the experimental data from reported in \cite{PNgabonziza_2020}, which are represent by the wide grey lines as described in the text. The theoretical calculations have been scaled by a factor of 5, which is similar to the normalization factors found in the previous works of $\sim$6 for Sr$_3$Ru$_2$O$_7$~\cite{MPAllan_2013} and  3-5 for Sr$_2$RuO$_4$~\cite{ATamai_2017}.\CBK
Comparison of theory and experiment for narrow bands near the Fermi energy in Sr$_4$Ru$_3$O$_{10}$ along the line from $\Gamma$ to $M$.   The experimental results from \cite{PNgabonziza_2020} are represented by open and grey solid circle curves as described in the text.  The theoretical bands are calculated with Hubbard $U = 1.5$ eV and they have been scaled by a factor of 5, which is similar to the renormalization factors found in the previous works of $\sim$6 for Sr$_3$Ru$_2$O$_7$~\cite{MPAllan_2013} and  3-5 for Sr$_2$RuO$_4$~\cite{ATamai_2017}.\CBK}
	\label{fig:flat-bands-comp}
\end{figure}

In addition to the Fermi surface map reproduced in Fig.~\ref{fig:fs-ARPES}, the experimental paper~\cite{PNgabonziza_2020} also reports bands around the zone center.  We can compare directly with the two bands along the $\Gamma$-M direction analyzed most carefully in ~\cite{PNgabonziza_2020}.  For measurements with Linear Horizontal Polarization (LHP) there is a band that curves upward with a band width of order ~40 meV; the dispersion shown in Figure 4c. of ~\cite{PNgabonziza_2020} is reproduced here as the open circle curve in Fig.~\ref{fig:flat-bands-comp}.  For measurements with Linear Vertical Polarization (LVP) there is a band that has a maximum at $\Gamma$ approximately 20 meV below the Fermi energy and curving downward;
%to ~100 meV roughly 1/2 the way to the zone boundary at the M point, 
the dispersion reported in Figure S2c in the supplementary material of~\cite{PNgabonziza_2020} is shown here as the grey solid circle curve in Fig.~\ref{fig:flat-bands-comp}.  

These results are compared with the theoretical calculations in Fig.~\ref{fig:flat-bands-comp}, where the energies of the calculation bands have been scaled by a factor of 5.  We have chosen to present the results for calculations with $U$ = 1.5 eV. The theoretical results for other values of U would be similar except the majority spin bands would be shifted upward or downward.
The renormalization factor of 5 is very reasonable compared with the previous works of factors $\sim$6 for Sr$_3$Ru$_2$O$_7$~\cite{MPAllan_2013} and  3-5 for Sr$_2$RuO$_4$~\cite{ATamai_2017}.

There are other features evident in the experimental ARPES results that clearly indicate intensity near the Fermi energy near the X point, labeled $\gamma$ in Fig.~\ref{fig:fs-ARPES} and described in the text of~\cite{PNgabonziza_2020} along the M to X lines.  These are also present in the calculated bands as indicated by the narrow bands along the $\Gamma$ to X and M to X lines which are in the same energy range as those shown in Fig.~\ref{fig:flat-bands-comp}.  However, these features are not presented in as much detail as the results shown in Fig.~\ref{fig:flat-bands-comp}. We do not attempt to analyze the narrow bands further. The theory is not at a stage where the bands can be predicted in detail and further work will need a combination of high resolution experiments in conjunction with theoretical calculations.  In addition, we should bear in mind various caveats:  the dispersion of the bands perpendicular to the layers,  the resolution reported in the paper is or order 20 meV, and there may be other effects like impurity scattering and surface states.
\CBK

\section{Consequences for other properties}
\label{sec:Consequences}

The most novel properties of Sr$_4$Ru$_3$O$_{10}$ relate to magnetism. In addition to the ferromagnetic moment along the $c$ axis there is metamagnetic behavior for applied magnetic fields parallel to the layers.  Such a large
susceptibility can be explained by narrow bands at the Fermi energy, which has been proposed as the explanation for the observed metamagnetic behavior in Sr$_4$Ru$_3$O$_{10}$~\cite{MKCrawford_2002,FWeickert_2017,YJJo_2007,GCao_2003,ZQMao_2006,DFobes_2010,YLiu_2010,YLiu_2010,ECarleschi_2014,WSchottenhamel_2016}
and studied in more detail in Sr$_3$Ru$_2$O$_{7}$~\cite{SAGrigera_2001,RABorzi_2007,ATamai_2008,DJSingh_2001,MPAllan_2013,JLee_2009,SAGrigera_2004,BBinz_2004}.
Narrow bands in the right energy range are found in the present work for DFT calculations using the PBESol functional and DFT+U for values of $U$ up to around $U=2.0$ eV. However, we have concluded that the states of the art for the theory and the experimental work done so far are not definitive enough to establish the nature of the narrow bands in detail.

A result of this work is that for Hubbard interaction $U>1.0$ eV, Sr$_4$Ru$_3$O$_{10}$  is predicted to be a half-metal, with only minority-spin bands at the Fermi energy.  Such materials are potentially of technological interest for spintronics applications~\cite{IZutic-RevModPhys.76.323} where the transport properties involve interesting many-body effects~\cite{MKatsnelson-RevModPhys.80.315}.
%A half-metal has been found in DFT calculations~\cite{PRivero_2017} using hybrid functionals which are in some ways analogous to DFT+U functionals.
The present calculations present interesting possibilities of a nearly-half-metal with almost filled majority spin bands that have large susceptibility due to narrow bands at the Fermi energy.
%Bands with this character result in the theoretical calculation for a range of $U$ values and appear to found in ARPES experiments~\cite{PNgabonziza_2020}.
This would combine effects related to metamagnetism and half-magnetism that could be controlled by applied fields. For larger interactions $U$, there would be no narrow bands near the Fermi energy and the calculations would find behavior more like other half-metals.
%
%There is a maximum spin moment, i.e., 3 up and 1 down electrons per Ru atom.  This is not a contradiction with the fact that the average magnetization does not correspond to fully polarized spins. Even if a mean field calculation predicts a full moment with integer spins, quantum fluctuations can lead to decreased average magnetization along a given direction in agreement with the experimental findings.

\section{Conclusions}
\label{sec:Conclusions}

The main conclusion of this work is the effect on the electronic bands of Sr$_4$Ru$_3$O$_{10}$ due to the ferromagnetic order.
% that the electronic band structure is very different from single layer Sr$_2$RuO$_{4}$ and double-layer Sr$_3$Ru$_2$O$_{7}$, even though the structure of Sr$_4$Ru$_3$O$_{10}$ is analogous to Sr$_3$Ru$_2$O$_{7}$.
Because the bands  are spin-polarized with a large ferromagnetic moment (known experimentally and found in the theoretical calculations), the majority spin bands are full or almost full and the minority spin bands have an occupancy of approximately 1/3.  The result is that  there are wide bands crossing the Fermi energy due to the minority spin, whereas there are narrow bands at or near the Fermi energy for the majority spin. When the spin-orbit coupling is included, there is still the distinction of two types of bands and there is an additional structure in the narrow bands near the Fermi energy. The fact that there are several narrow bands leads to a large density of states that varies rapidly near the Fermi energy, which is a way to understand metamagnetism as well as more than one metamagnetic transition as has been observed in Sr$_4$Ru$_3$O$_{10}$ \cite{ECarleschi_2014}.

The experimental ARPES measurements~\cite{PNgabonziza_2020} are in good agreement with the theoretical predictions for spin-polarized bands and qualitatively different from the calculated bands if the systems is constrained to be non-spin-polarized.\CBK As discussed in Sec.~\ref{sec:Exp-comparison}, the wide bands for the minority spins explain the dominant features of the ARPES data that are roughly square and circular Fermi surfaces, whereas the narrow bands of the majority spins are in qualitative agreement with the observations of narrow bands around the zone center. The features due to the minority spins remain the same in all the spin-polarized calculations. However, the position of the majority spin bands relative to the Fermi energy is very sensitive to the choice of the functional, which is illustrated by the large changes as a function of an added Hubbard $U$ interaction on the Ru $d$ states.
%Therefore, we conclude that a complete understanding of the electronic properties will require additional theoretical calculations in concert with experimental studies with higher resolution than those done up to now.

%\CRD OMITTED REPETITIVE PARTS \CBK

%We have also calculated the spectra with the system constrained to be non-polarized.  The $t_{2g}$ bands are 2/3 full, so that the $t_{2g}$ Fermi energy falls in a range with many bands.  The Fermi surface is very different and this solution can be ruled out by the comparison with the ARPES data.  It is interesting to note that the non-spin-polarized bands can be understood as the triple-layer analogue of Sr$_2$RuO$_{4}$ and Sr$_3$Ru$_2$O$_{7}$, so that the difference shows clearly the large effect of ferromagnetism.

Finally, it is important to put the present work in perspective in the big picture of electronic interactions and correlation in the layered strontium ruthenates and, more generally, transition metal oxides.  The ARPES measurements indicate narrow bands roughly a factor of 5 narrower than those found in the present DFT calculations.  This is comparable to the other layered strontium ruthenates ~\cite{MPAllan_2013, ATamai_2017}
and indicates similar effects of correlations.  In addition, however, there is another effect of interactions, the magnitude of the spin-splitting in Sr$_4$Ru$_3$O$_{10}$ due to the ferromagnetic order.  Our results show that there are narrow majority-spin bands for a range of added interactions $U$ on the Ru $d$ states, which appears to be qualitatively consistent with the experimental data. But our knowledge of functionals and the energy resolution of the ARPES experiments done so far are not sufficient to pin down details. Exploring the nature of the narrow bands in future work can provide insight into the role of interactions and correlations in the entire range of these layered systems and transition metal oxides in general.

\section{Acknowledgements}
\label{sec:Acknowledgements}

This work was supported by the 2019 Innovation Fund
of the American Physical Society though the US-Africa Initiative in
Electronic Structure and by the Abdus Salam International Centre for Theoretical Physics.
We gratefully acknowledge the African School for Electronic Structure Methods and Applications (ASESMA) which was the stimulus for this work, and we thank Emanuela Carleschi, Johan Chang, Mario Cuoco, Bryan P. Doyle, and Masafumi Horio for
fruitful discussions.
This work used the Extreme Science and Engineering Discovery Environment (XSEDE), which is supported by National Science Foundation grant number ACI-1548562, and the Centre for High Performance Computing (CHPC),
South Africa.
Finally, we are indebted to Alexei Fodorov, Jonathan Denlinger and James Allen for bringing to our attention unpublished ARPES measurements on Sr$_4$Ru$_3$O$_{10}$ referred to in the footnote, reference 53.
\CBK

\bibliography{Bibfile-SrRuO}
%\bibliography{Bibfile-SrRuO_v3}
%\bibliography{ELSGP,RMM-68-86-articles,BIBFILE-COMBINED-2017,NEW-REFS-2017}

%\end{multicols}
\end{document}